\documentclass[lettersize,journal]{IEEEtran}
\usepackage{amsmath,amsfonts}
\usepackage{algorithmicx}
\usepackage{algpseudocode}
\usepackage{algorithm}
\usepackage{array}
\usepackage{textcomp}
\usepackage{stfloats}
\usepackage{url}
\usepackage{verbatim}
\usepackage{graphicx}
\usepackage{wrapfig}
\usepackage{amssymb}

\usepackage[numbers]{natbib}

\usepackage[utf8]{inputenc} %
\usepackage[T1]{fontenc}    %
\usepackage{url}            %
\usepackage{booktabs}       %
\usepackage{amsfonts}       %
\usepackage{nicefrac}       %
\usepackage{microtype}      %

\usepackage{color,xcolor}
\usepackage{epsfig}
\usepackage{graphicx}
\usepackage{subfiles}

\usepackage{adjustbox}
\usepackage{array}
\usepackage{booktabs}
\usepackage{colortbl}
\usepackage{float,wrapfig}
\usepackage{hhline}
\usepackage{multirow}
\usepackage{subcaption} %
\usepackage[labelfont={bf},labelsep={period},font={small}]{caption}

\let\llncssubparagraph\subparagraph
\let\subparagraph\paragraph
\usepackage[compact]{titlesec}
\let\subparagraph\llncssubparagraph

\usepackage{amsmath,amsfonts,amssymb}
\usepackage{bm}
\usepackage{nicefrac}
\usepackage{microtype}

\usepackage{changepage}
\usepackage{extramarks}
\usepackage{fancyhdr}
\usepackage{lastpage}
\usepackage{setspace}
\usepackage{soul}
\usepackage{xspace}

\usepackage{enumerate}

\usepackage{times}

\usepackage{pbox}
\usepackage{footnote}
\usepackage{tablefootnote}

\usepackage{paralist}

\usepackage{enumitem}
\setitemize{noitemsep,topsep=0pt,parsep=0pt,partopsep=0pt}

\definecolor{mydarkgreen}{rgb}{0.02,0.6,0.02}

\usepackage{tabu}
\usepackage{pifont}
\usepackage[normalem]{ulem}
\usepackage{nccmath}
\usepackage{dsfont}
\usepackage{makecell}
\usepackage{scrextend}
\usepackage{hyperref}

\usepackage{pifont}
\newcommand{\xmark}{\ding{55}} %
\hypersetup{colorlinks,
    linkcolor=red,citecolor=citecolor,urlcolor=blue}

\definecolor{citecolor}{RGB}{34,139,34}
\definecolor{mydarkblue}{rgb}{0,0.08,1}
\definecolor{mydarkgreen}{rgb}{0.02,0.6,0.02}
\definecolor{mydarkred}{rgb}{0.8,0.02,0.02}
\definecolor{mydarkorange}{rgb}{0.40,0.2,0.02}
\definecolor{mypurple}{RGB}{111,0,255}
\definecolor{myred}{rgb}{1.0,0.0,0.0}
\definecolor{mygold}{rgb}{0.75,0.6,0.12}
\definecolor{myblue}{rgb}{0,0.2,0.8}
\definecolor{mydarkgray}{rgb}{0.,0.2,0.2}

\definecolor{lightred}{RGB}{255,235,235}
\definecolor{lightgreen}{RGB}{235,255,235}
\definecolor{lightblue}{RGB}{235,235,255}
\definecolor{lightcyan}{RGB}{235,255,255}
\definecolor{lightmagenta}{RGB}{255,235,255}
\definecolor{lightyellow}{RGB}{255,255,235}

\definecolor{qxkcolor}{RGB}{215,235,255}
\definecolor{softmaxcolor}{RGB}{230,235,255}
\definecolor{probxvcolor}{RGB}{255,255,235}

\definecolor{topkcolor}{RGB}{255,235,235}
\definecolor{zecolor}{RGB}{255,255,235}
\definecolor{dynacolor}{RGB}{235,255,255}

\definecolor{reviewcolor}{RGB}{0,0,200}

\newcommand{\ignorethis}[1]{}

\renewcommand*{\thefootnote}{\fnsymbol{footnote}}

\makeatletter
\DeclareRobustCommand\onedot{\futurelet\@let@token\@onedot}
\def\@onedot{\ifx\@let@token.\else.\null\fi\xspace}

\makeatother

\makeatletter
\newcommand\footnoteref[1]{\protected@xdef\@thefnmark{\ref{#1}}\@footnotemark}
\makeatother

\definecolor{mydarkblue}{rgb}{0,0.08,1}

\definecolor{mydarkred}{rgb}{0.8,0.02,0.02}
\definecolor{mydarkorange}{rgb}{0.40,0.2,0.02}
\definecolor{mypurple}{RGB}{111,0,255}
\definecolor{myred}{rgb}{1.0,0.0,0.0}
\definecolor{mygold}{rgb}{0.75,0.6,0.12}
\definecolor{mydarkgray}{rgb}{0.66, 0.66, 0.66}
\definecolor{mygray}{gray}{0.9}
\definecolor{bigaired}{RGB}{156, 0, 0}
\definecolor{uclablue}{RGB}{39, 116, 174}

\newcommand{\blfootnote}[1]{%
  \begingroup
  \renewcommand\thefootnote{}\footnote{#1}%
  \addtocounter{footnote}{-1}%
  \endgroup
}

\IEEEoverridecommandlockouts
\begin{document}

\title{Harnessing Photonics for Machine Intelligence}

\author{
Hanqing Zhu$^{1*}$,
Shupeng Ning$^{1*}$,
Hongjian Zhou$^{2*}$,
Ziang Yin$^{2*}$,
\\
Ray T. Chen$^1$,
Jiaqi Gu$^{2\dagger}$,
David Z. Pan$^{1\dagger}$,
\\$^1$The University of Texas at Austin\quad 
$^2$Arizona State University
}

\maketitle
\bstctlcite{IEEEexample:BSTcontrol} 

\begin{abstract}
The exponential growth of machine-intelligence workloads is colliding with the power, memory, and interconnect limits of the post-Moore era, motivating compute substrates that scale beyond transistor density alone. 
Integrated photonics is emerging as a candidate for artificial intelligence (AI) acceleration by exploiting optical bandwidth and parallelism to reshape data movement and computation.
This review reframes photonic computing from a circuits-and-systems perspective, moving beyond building-block progress toward cross-layer system analysis and full-stack design automation.
We synthesize recent advances through a bottleneck-driven taxonomy that delineates the operating regimes and scaling trends where photonics can deliver end-to-end sustained benefits.
A central theme is cross-layer co-design and workload-adaptive programmability to sustain high efficiency and versatility across evolving application domains at scale.
We further argue that Electronic-Photonic Design Automation (EPDA) will be pivotal, enabling closed-loop co-optimization across simulation, inverse design, system modeling, and physical implementation.
By charting a roadmap from laboratory prototypes to scalable, reproducible electronic-photonic ecosystems, this review aims to guide the CAS community toward an automated, system-centric era of photonic machine intelligence\blfootnote{$^*$Equal contributions; $^\dagger$Corresponding authors}.

\end{abstract}

\begin{IEEEkeywords}
Photonic computing, AI hardware, cross-layer co-design, electronic-photonic design automation.
\end{IEEEkeywords}

\section{Reframing Photonics for AI Compute: An Emerging Substrate}
\label{sec:intro}

Machine intelligence~\cite{lecun2015deep} has become the defining workload of modern computing, evolving from early deep neural networks that excelled at pattern recognition~\cite{krizhevsky2012imagenet} to today’s foundation models. 
Underpinned by empirical scaling laws~\cite{kaplan2020scaling,hoffmann2022training}, this evolution has shifted the field from isolated algorithmic novelty to systematic scaling of parameters and data, a trajectory that demands vastly more compute.
Crucially, this paradigm now extends beyond training to inference: test-time scaling (TTS)~\cite{wei2022chain,wu2025inference,cong2025can} unlocks stronger capabilities by allocating additional computation at deployment, a trend accelerated by reasoning-centric models~\cite{jaech2024openai,guo2025deepseek}. 
This shift fundamentally alters the computational landscape: inference costs are transitioning from being linearly proportional to model size to scaling exponentially with the complexity of reasoning and agentic tasks~\cite{jaech2024openai, wu2025inference}.
Consequently, compute availability has emerged as the central determinant of the "supply of intelligence," creating an urgent need to ensure that hardware performance does not become the bottleneck for the pace of machine intelligence.

For decades, the industry relied on Moore’s Law~\cite{moore1998cramming} and Dennard scaling~\cite{bohr200930} to deliver near-automatic performance gains to meet the rising compute demand.
However, this roadmap has now collided with fundamental physical limits in the single-digit-nanometer regime~\cite{waldrop2016more, fang2019towards}. As voltage scaling stagnates and quantum effects (e.g., tunneling) intensify, energy efficiency has failed to keep pace with density, making power, not transistor count, the dominant system limiter~\cite{horowitz2014computing}. 
This results in the era of \emph{dark silicon}~\cite{esmaeilzadeh2011dark}, where thermal constraints prevent fully utilizing the available hardware. Consequently, transistor scaling alone is structurally incapable of sustaining the exponential growth of artificial intelligence (AI) workloads~\cite{shalf2020future}.
This breakdown motivates alternative computing substrates and architectures that can deliver scalable compute to support the growing intelligence.

While photonics has established itself as the definitive solution to the data-movement wall in interconnects~\cite{wan2025integrating}, its utility is now expanding into the domain of computation itself~\cite{mcmahon2023physics, xu2023integrated,ning2024photonic}.
We are witnessing a \textit{resurgence of interest in optical computing}, distinguished by a fundamental strategic shift: rather than mimicking general-purpose digital logic, an approach demanding challenging cascadability and signal restoration~\cite{miller2010optical}, the community is increasingly targeting \textit{special-purpose, predominantly analog} accelerators~\cite{mcmahon2023physics, solli2015analog, ahmed2025universal}.

This pivot harnesses the intrinsic properties of light, such as high bandwidth, low latency, and massive parallelism, to perform efficient linear transformations, a capability that aligns precisely with the workload of modern deep learning.
Since foundation models are dominated by dense Matrix-Vector Multiplication (MVM) yet exhibit remarkable tolerance to low-precision computations~\cite{gholami2021survey, dettmers2022gpt3},
they are ideally suited to the analog domain. 
This algorithmic robustness allows optical cores to serve as \textit{high-throughput, specialized primitives for the next generation of machine intelligence}.

Driven by this promise, recent years have witnessed rapid progress in optical computing prototypes, particularly photonic integrated circuit (PIC)-based optical neural networks (ONNs). This evolution is reinforced by massive physical gains: recent milestones include 3.8~TOPS throughput~\cite{xu202111}, sub-femtojoule energy efficiency~\cite{heni2019plasmonic}, and the emerging realization of universal AI acceleration on photonic hardware~\cite{ahmed2025universal}.
While the field has predominantly demonstrated standard architectures, spanning multi-layer perceptron (MLP)~\cite{shen2017deep}, convolutional neural networks (CNNs)~\cite{xu202111,xu2024large}, and spiking NNs (SNNs)~\cite{chakraborty2019photonic,feldmann2019all}, it is now actively expanding toward \emph{advanced} AI workloads, e.g., Transformer-style architectures, the foundation-model primitive. Notably, \cite{zhu2024lightening} presents one of the first explicit mappings of Transformer computation patterns onto a photonic substrate, redesigning the architecture to accommodate the distinct structure of attention operators

Yet, despite these compelling prototypes, it remains unclear \textit{when} and \textit{how} optical computing delivers a sustained \emph{system-level} advantage. Prior surveys and many reported demonstrations emphasize isolated device- and circuit-level innovations, while system-critical factors, including workload mapping, data orchestration, memory hierarchy, and the non-trivial overheads of mixed-signal interfaces and control, are often simplified or omitted. 
As a result, reported ``optical-core'' metrics do not directly translate into deployable performance, and comparisons across architectures and workloads remain difficult to interpret.

To close this gap, Section~\ref{sec:related} establishes a \emph{system-level benchmarking} baseline using cross-layer simulation (e.g., our \textsc{SimPhony} framework~\cite{yin2024simphony}). By modeling the full heterogeneous electronic-photonic datapath, from photonic tensor cores (PTCs) to digital-to-analog (DAC)/analog-to-digital(ADC) converters, memory, and laser power, we quantify where photonics is competitive today under realistic assumptions. This benchmarking lens makes the ``system tax'' explicit and enables fair comparisons across representative PTC families and NN operators (e.g., linear layers and attention mechanisms).

Beyond point comparisons, however, a benchmark alone does not answer the forward-looking question: \textit{what must improve for photonic computing to scale with rapidly evolving AI workloads?} Section~\ref{sec:photonic_ai} performs a \emph{scaling analysis} to extract the critical dimensions that govern end-to-end efficiency, including area, parallelism (spatial/spectral/temporal), bit precision, and the degree to which interface costs can be amortized by reuse. Guided by these simulation-derived ``pressure points,'' we organize the literature into a bottleneck-driven taxonomy, highlighting how recent efforts aim to scale photonic systems along the dimensions that matter most at the full system level.

Finally, scaling photonic AI from prototypes to deployable heterogeneous EPICs requires \emph{full-lifecycle electronic-photonic design automation (EPDA)}.
Section~\ref{sec:epda} reviews emerging EPDA capabilities across design stacks, including AI-assisted device simulation and inverse design, photonic-electronic circuit-level co-simulation, system-level modeling, and layout synthesis automation.
We identify the remaining gaps and outline an EPDA roadmap toward future infrastructures to enable scalable, reproducible, and efficient design of photonic AI systems.

\textbf{Roadmap and Organization}. This survey arrives at a critical inflection point, as traditional electronic scaling struggles to meet the exponential demand for compute while integrated photonics matures into a viable heterogeneous accelerator. Unlike prior surveys that emphasize either device physics or isolated circuit techniques, we adopt a system-scalability lens: examining how photonics alters the computing landscape, when it provides a distinct advantage, what limits it, and what toolchains are required to make it practical.

The remainder of this paper is organized as follows:
\begin{itemize}
  \item \textbf{Section~\ref{sec:related}} establishes a system-level benchmarking baseline using cross-layer simulation, quantifying the regimes where photonics is competitive once mixed-signal interfaces, memory traffic, and optical link-budget constraints are included.
  \item \textbf{Section~\ref{sec:photonic_ai}} extracts the critical scaling dimensions that govern end-to-end performance and uses these simulation-derived pressure points to categorize photonic accelerator efforts into a bottleneck-driven taxonomy.
  \item \textbf{Section~\ref{sec:epda}} reviews the EPDA stack required to scale photonic AI from prototypes to deployable EPICs, including modeling, verification, physical design, and calibration-aware co-simulation.
\end{itemize}

\section{Quantifying Photonic Advantage: From Physical Promise to System-level Benchmark}
\label{sec:related}

While the intrinsic physical advantages of optical computing, such as high bandwidth and low latency, are well understood, translating these attributes into deployable AI performance requires rigorously accounting for system-level constraints. In realistic hybrid architectures, the optical core never operates in isolation; its realizable throughput and efficiency are fundamentally governed by both the photonic devices and the electronic periphery, specifically the data converters (ADCs/DACs) and memory interfaces, required to sustain them.

In this section, we deconstruct the photonic advantage by tracing the signal chain from physics to system layers: \begin{itemize} 
\item \textbf{Section~\ref{subsec:physics} (The Physics):} 
We revisit the \textbf{physical primitives}, isolating the specific properties of light (e.g., parallelism and passivity) that make it structurally efficient for linear algebra. 
\item \textbf{Section~\ref{subsec:simulation} (The System):} 
We conduct \textbf{system-level modeling and benchmarking}. By modeling the full datapath including mixed-signal overheads, we quantify photonic advantages and identify key system bottlenecks.
\end{itemize}

\subsection{Physical Primitives: Why Light Can Be Efficient}
\label{subsec:physics}

Despite the rapid evolution of architectures, from CNNs to Transformers, the underlying computational backbone remains invariant. 
Whether computing convolutions, linear projections, or attention scores, the vast majority of modern AI inference relies on a single fundamental primitive: \textbf{Matrix-Vector Multiplication}.
Below, we outline the primary physical attributes of light that allow it to execute these dense linear transformations with ultra-low latency, massive parallelism, and energy efficiency, beyond what is readily achievable with conventional electronics.

\begin{figure}[t]
    \centering
    \includegraphics[width=\columnwidth]{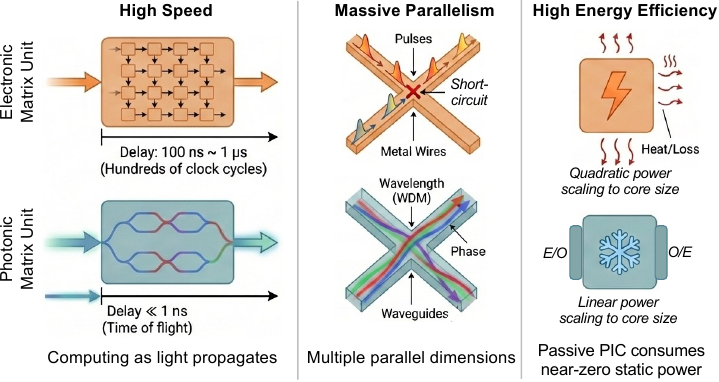}
    \caption{Physical advantages of photonic computing: ultra-low latency via RC-free propagation, high bandwidth density via multiplexing, and superior energy efficiency.
    }
    \label{fig:PhotonicAdv}
\end{figure}

\subsubsection{Low Latency via RC-Free Optical Propagation}

A common \textit{misconception} is that optical computing is faster simply because light travels ``faster.'' In practice, the group velocity in silicon/silicon-nitride waveguides is typically $0.3$--$0.5c$, comparable to propagation in well-designed electrical transmission lines~\cite{thevenaz2008slow,pozar2021microwave}. The advantage is instead \emph{how} delay scales with connectivity.

In dense CMOS interconnects, long wires and high fanout incur resistive-capacitive (RC) delays; the effective delay grows superlinearly with wire length in distributed RC regimes, and mitigating this requires repeater insertion, which adds area and power~\cite{weste2015cmos,eyerman2011fine}. 
Optical waveguides, in contrast, are "RC-free", i.e., the delay depends almost exclusively on the geometric path length (scaling linearly) and is effectively independent of capacitive loading. This enables optical signals to traverse centimeter-scale chips in sub-hundred-picosecond flight times, offering a critical latency advantage for the high-fanout broadcasting and global data distribution required by modern neural networks.

\subsubsection{High Bandwidth Density via Multiplexed Channels}

A frequent critique of photonics is the ``density gap''.
In modern CMOS, transistors scale to $\sim\!\! 10^{10}\,\mathrm{cm}^{-2}$, while fabricated photonic components are typically orders of magnitude larger due to the diffraction limit~\cite{mcmahon2023physics}. However, \textit{raw component density is not the right proxy for throughput}. 
In reality, the "truth" of photonic scaling lies in its ability to exploit dimensions of parallelism inaccessible to electronics.

For optical communication and computing, information is encoded onto electromagnetic waves that propagate through fibers or on-chip waveguides at carrier frequencies in the terahertz-petahertz range.
Since they are not constrained by RC delay and Joule heating that confine practical electrical signaling rates to at most a few gigahertz (GHz)~\cite{ng2022power,eyerman2011fine}, optical signals inherently afford a wider usable bandwidth.
More importantly, the neutrality and bosonic nature of photons allow multiple optical modes to coexist within a shared waveguide without mutual exclusion or Coulombic repulsion~\cite{griffiths2018introduction,saleh2019fundamentals}. 
Orthogonal channels such as wavelength-division, polarization, spatial modes, and temporal encoding can be densely multiplexed with negligible crosstalk, yielding massive parallelism within a bulky photonic footprint~\cite{saleh2019fundamentals,rizzo2023massively}.

This multiplexing capability maps naturally onto the intrinsic fan-in and fan-out patterns of MVMs, alleviating the routing congestion, signal interference, and limited parallelism that increasingly constrain advanced CMOS platforms.

\subsubsection{High Energy Efficiency via Linear Power Scaling} %
Photonic computing can offer superior energy efficiency by avoiding the resistive heating that fundamentally limits electronic AI accelerators. 
In CMOS technologies, dynamic power scales as $P_{\mathrm{dyn}} \propto C V_{dd}^{2} f$, and increases in operating frequency $f$ typically require proportional increases in supply voltage $V_{dd}$ to maintain switching speed. 
This voltage-frequency coupling leads to superlinear power scaling at high-performance operating points~\cite{ng2022power}. 
Additional frequency-dependent losses, such as skin-effect-induced current crowding in metal interconnects, further exacerbate ohmic heating~\cite{pozar2021microwave}. 

By leveraging capacitive electro-optic mechanisms, EPIC achieves high modulation speeds while maintaining near-zero static power, avoiding the resistive leakage and thermal dissipation inherent to active electronic transistors. Since the core matrix operations are subsequently performed via passive wave propagation, the dynamic power consumption is fundamentally decoupled from the computational complexity of the matrix interior. Unlike electronic processors, which suffer from quadratic energy growth with matrix size dictated by active switching for every arithmetic operation, EPIC confines energy expenditure to the I/O interfaces, thereby exhibiting a linear scaling trajectory with both operating frequency and input dimension. This advantage is further strengthened by passive photonic paradigms, including architectures based on phase-change materials (PCM)~\cite{feldmann2021parallel}, diffractive structures~\cite{zhu2022space,xu2024large}, and metasurfaces~\cite{wang2022integrated}, which reduce electrical overhead by implementing linear transforms directly via propagation.

\subsection{System-Level Benchmark: A Cross-Layer Simulation via SimPhony~\cite{yin2024simphony}}
\label{subsec:simulation}

\begin{figure*}
    \centering
    \includegraphics[width=0.9\textwidth]{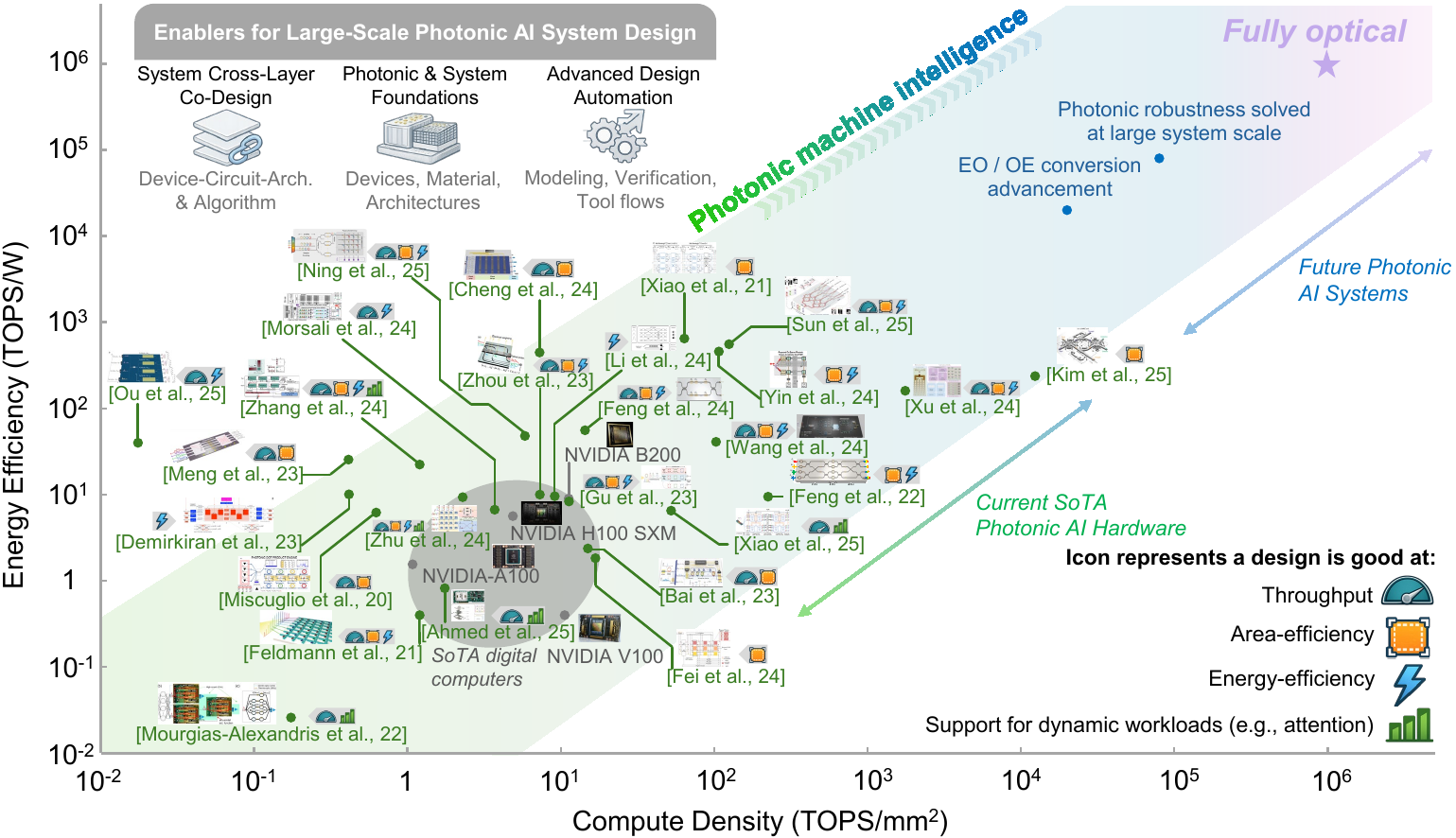}
    \caption{System-level energy efficiency vs. compute density comparison between silicon-photonic accelerators and digital GPUs. 
    The axes report effective energy efficiency (TOPS/W) and compute density (TOPS/mm$^2$) under reported or derived system-level assumptions.
    GPU points correspond to NVIDIA V100~\cite{nvidiaV100Datasheet2020}, A100~\cite{nvidiaA100Datasheet2022}, H100 SXM~\cite{nvidiaH100Datasheet2023}, and B200~\cite{nvidiaHGXB200PCFSummary2025}, using vendor-reported peak INT8 performance and die area.
    Photonic points are compiled from prior silicon-photonic accelerator works~\cite{ahmed2025universal, shen2017deep, xu2024large, zhu2024lightening, feldmann2021parallel, tait2016microring, 10.1063/5.0001942, feng2022compact, ning2025hardware, meng_tempo, yin_dac2025, Demirkiran2023, 9820779, 10.1063/5.0255883, 10.1145/3658617.3697706, a557a39c698546f1902a12c137b1913f, Wang2024, 10.1145/3650200.3656609, Sun2025, 10.1063/5.0070913, Zhou2023, ou2025hypermultiplexed, 202501995, feng2024integrated,meng2023compact,cheng2024multimodal} and are annotated to reflect which scaling dimensions each design emphasizes (area efficiency, energy efficiency, robustness at scale, and dynamic-workload capability, such as attention). 
    }
    \label{fig:arch_efficiency}
\end{figure*}

While photonic primitives offer intrinsic physical advantages, deployable utility is ultimately determined by the end-to-end \emph{system tax} imposed by mixed-signal interfaces and memory, rather than isolated device metrics. 

From an architecture standpoint, two metrics are especially diagnostic:
\begin{itemize}
    \item \textbf{Effective Energy Efficiency (TOPS/W):} end-to-end throughput per wall-plug power, governing thermal feasibility and operating cost at scale.
    \item \textbf{Compute Density (TOPS/mm$^2$):} sustained throughput per integrated area, reflecting how much performance can be delivered per chip footprint once peripheral overheads are included.
\end{itemize}

Figure~\ref{fig:arch_efficiency} maps representative photonic prototypes against state-of-the-art graphics processing units (GPUs) on the efficiency-density plane. 
The plot highlights rapid progress: successive demonstrations are pushing toward the upper-right, and several photonic tensor-core designs report regimes of tera operations per second per Watt (TOPS/W) and/or TOPS/mm$^2$ that exceed typical GPU operating points.

However, the reported photonic points span orders of magnitude, and even superficially similar photonic approaches can appear far apart.
This spread is primarily driven by non-unified evaluation processes and assumptions, workload choice, precision and utilization assumptions, and whether key system taxes are consistently included (e.g., laser power under realistic insertion loss, ADC/DAC energy at the required bandwidth, and the control/calibration overheads needed to maintain operating points). Without a unified benchmarking methodology, it is difficult to isolate true bottlenecks or predict how architectural scaling will translate into deployable advantage.

\subsubsection{Simulation Tool: a Cross-Layer Modeling Framework}

\begin{figure}
    \centering
    \vspace{-3pt}
    \includegraphics[width=\columnwidth]{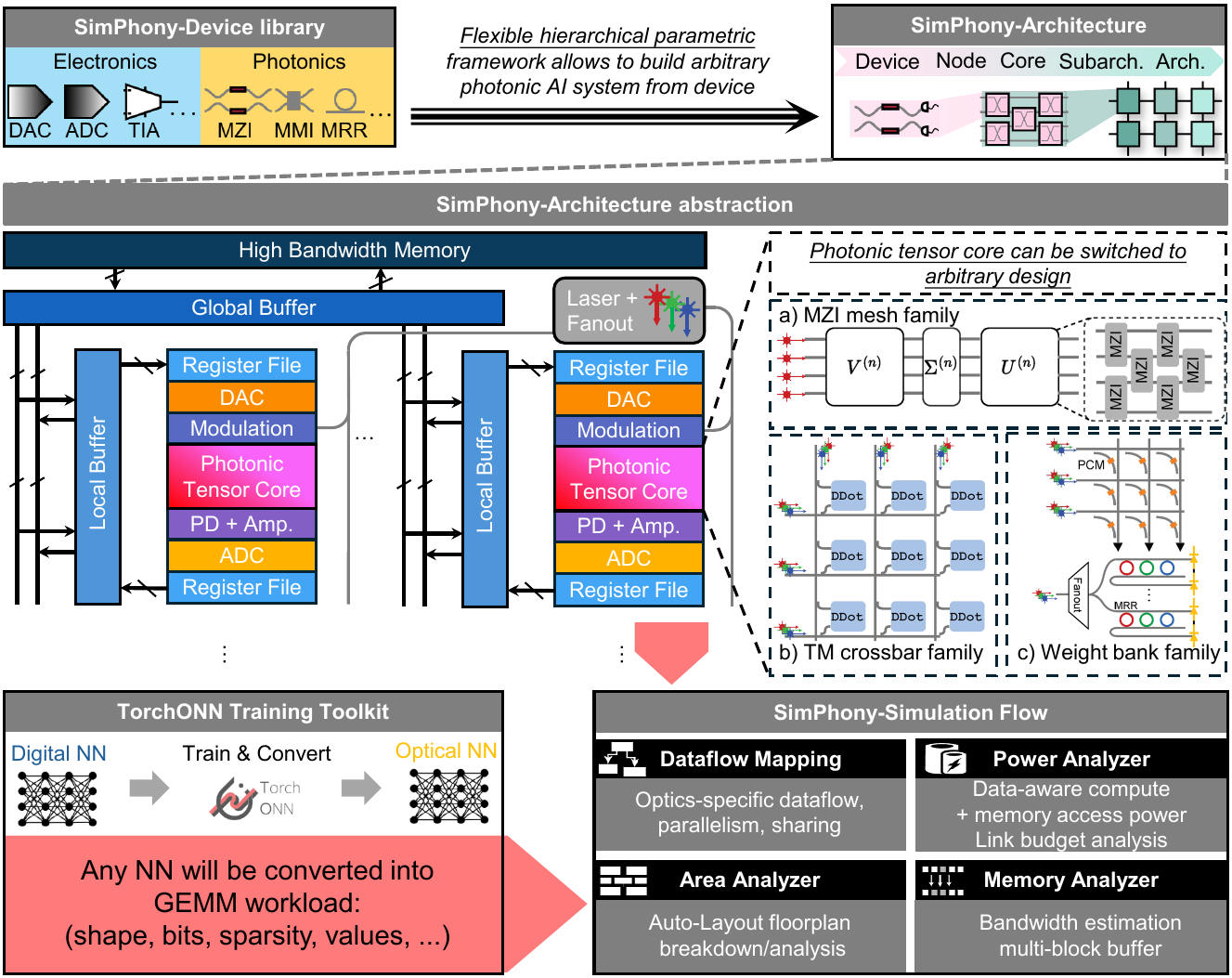}
    \vspace{-10pt}
    \caption{\textsc{SimPhony}~\cite{yin2024simphony} cross-layer modeling framework for heterogeneous electronic-photonic AI systems. 
    A modular PTC can be instantiated from different architecture families under a unified system interface. 
    The framework maps NN operators into GEMM workloads, generates optics-specific dataflow, and evaluates end-to-end metrics using analyzers for memory traffic, area/floorplan, power/energy breakdown, and optical link budget.
    }
    \label{fig:Archabstraction}
    \vspace{-10pt}
\end{figure}

To benchmark photonic AI accelerators under realistic constraints, we leverage \textsc{SimPhony}~\cite{yin2024simphony}, a cross-layer modeling framework capable of simulating heterogeneous electronic-photonic systems from the device level up to architecture. Rather than relying on isolated figures of merit for the photonic core, \textsc{SimPhony} evaluates end-to-end performance by explicitly modeling the full signal chain: the photonic compute fabric, the mixed-signal interfaces (drivers, modulators, trans-impedance amplifiers (TIAs), ADCs/DACs), and memory required to sustain high-bandwidth operation.

As illustrated in Fig.~\ref{fig:Archabstraction}, the framework composes a parametric system model from physical device and circuit building blocks. It then (i) generates optics-specific dataflows exploiting wavelength, time, and spatial parallelism; (ii) tracks memory traffic through multi-level buffers and off-chip memory; (iii) enforces optical link-budget constraints to determine the required laser power; and (iv) produces layout-aware area estimates and a granular energy breakdown (laser, modulation, readout, conversion, and memory). This holistic accounting is essential because while many photonic architectures exhibit favorable \emph{core-level} metrics, their deployable efficiency is strictly governed by "system taxes" that dominate at scale:

\begin{itemize}
    \item \textbf{Memory Delivery and Utilization:} Bandwidth limits and the energy cost of data movement can throttle the optical core, often outweighing compute energy.
    \item \textbf{Optical Loss $\rightarrow$ Laser Power:} Critical-path insertion loss and signal-to-noise ratio (SNR) targets directly dictate wall-plug laser power:
    \begin{equation}
        \label{eq:LaserPower}
        \small
        P_{laser} = \frac{10^{(S_{req} + IL) / 10} \cdot 2^{b_{out}}}{\eta_{WPE}} \cdot \frac{1}{1 - 0.1^{ER / 10}},
    \end{equation}
    where $IL$ is the total insertion loss, $S_{req}$ is the required SNR margin, $b_{out}$ is the effective output precision, $\eta_{WPE}$ is the laser wall-plug efficiency, and $ER$ is the modulator extinction ratio.
    
    \item \textbf{EO/OE Interfaces:} Driver and converter energy scales linearly with sampling rate and exponentially with resolution. We parameterize this using Walden-FoM scaling:
    \begin{equation}
        \begin{aligned}
            P_{DAC}(b_{in},f) &= FoM_{DAC}\cdot2^{b_{in}}\cdot f\\
            P_{ADC}(b_{out},f) &= FoM_{ADC}\cdot2^{b_{out}}\cdot f.
            \label{eq:dac_power}
        \end{aligned}
    \end{equation}
    This model highlights a recurring system-level reality: unless conversion is aggressively amortized (via reuse, reduced precision, or lower rates), the DAC/ADC and modulation overheads will dominate the total energy.
    
    \item \textbf{Photonics-Aware Mapping:} Parallelism yields system benefits only if the dataflow effectively amortizes conversion and movement costs, rather than simply multiplying the number of required interfaces.
\end{itemize}

\subsubsection{Simulation Setup: Photonic Tensor Core and Workload}

\noindent
\textbf{The architecture-workload interaction.}~
System-level energy and throughput are not intrinsic properties of a photonic core; rather, they emerge from the interaction between the \emph{tensor core topology} (specifically, how it encodes weights) and the \emph{workload characteristics} (specifically, how frequently weights change). 
This coupling is critical in the modern AI landscape, which is transitioning from static-weight workloads (e.g., CNNs) to dynamic, weight-free mechanisms (e.g., Attention) that require frequent operand refreshing. 
To capture this spectrum, we select representative architectures and workloads that span the design space from high-reuse static execution to reuse-free dynamic execution.

\noindent
\textbf{Representative PTC families.}~
We categorize the diverse landscape of photonic AI systems into three families based on their physical weight-encoding mechanisms. We select one representative design for each to be evaluated under identical conditions ($8\times8$ core, 12 wavelengths, 8-bit, 5 GHz):

\begin{itemize} 
    \item \ding{202} \textbf{\emph{Mach-Zehnder Interferometer (MZI) Mesh Family (Coherent/Static-only):}} Exemplified by Clements-style arrays~\cite{shen2017deep, Demirkiran2023}, this family represents the coherent computing paradigm. It relies on unitary transformations and typically assumes slowly tunable phase shifters, making it highly efficient for static weights but challenging for rapid reconfiguration. We use the \textit{coherent nanophotonic circuit}~\cite{shen2017deep} as the exemplar.
    
    \item \ding{203} \textbf{\emph{Weight-Bank Family (Incoherent/Both dynamic and static):}} Typified by broadcast-and-weight architectures like microring resonator (MRR) banks~\cite{tait2017neuromorphic} and PCM arrays~\cite{feldmann2021parallel}. These designs offer high area density but, like MZI meshes, generally leverage static weight reuse to amortize the cost of precise thermal/PCM tuning. We use the \textit{MRR weight bank}~\cite{tait2016microring} as the representative design.
    
    \item \ding{204} \textbf{\emph{Time-Multiplexed Crossbar Family (Both dynamic and static):}} Represented by engines like Lightening-Transformer~\cite{zhu2024lightening} and TeMPO~\cite{meng_tempo}. These architectures are explicitly designed for dynamic dataflows; they employ high-speed modulators for all operands and exploit time-multiplexing to minimize interface overheads during rapid weight updates. We use \textit{Lightening-Transformer}~\cite{zhu2024lightening} as the representative design.
\end{itemize}

\noindent
\textbf{Workloads: static linear vs.\ dynamic attention.}~
Early optical accelerators largely targeted CNN/MLP-style layers where a \emph{trained weight matrix is reused} across many input activations, making weight-stationary mappings and slower/low-power weight-tuning mechanisms plausible.
In contrast, modern Transformers are dominated by \emph{attention micro-kernels} whose heaviest general matrix multiplications (GEMMs) include \emph{activation-activation} products (e.g., $QK^{\mathsf{T}}$), where both operands are token-dependent at runtime.
This reduces reuse and stresses the very subsystems that often dominate the system tax, conversion bandwidth, modulation/update rate, and memory traffic, thereby invalidating assumptions that are benign under static layers.
To capture this shift with a fair, compute-matched comparison, we evaluate two GEMM-equivalent operators with identical multiply-accumulate (MAC) counts:

\begin{itemize}
    \item \textbf{Dynamic (attention):}
    a representative self-attention micro-kernel with sequence length $S=1024$ and embedding dimension $D=512$, focusing on the $QK^{\mathsf{T}}$ product.
    Concretely, we model a batch of 4 query vectors multiplying a key matrix of width $S$, yielding 4$\times$512$\times$1024 MACs.

    \item \textbf{Static (linear):}
    a compute-matched batch-$4$ linear projection of shape 512$\rightarrow$1024, which also requires 4$\times$512$\times$1024 MACs.
    This serves as the weight-stationary baseline (representative of linear/conv projections after lowering) to contrast against attention-style dynamics.
\end{itemize}

\begin{figure}
    \centering
    \includegraphics[width=\columnwidth]{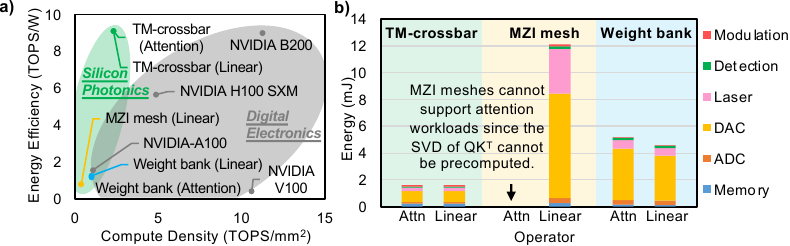}
    \caption{
    (a) Energy efficiency and compute density comparison between photonic and digital electronic hardware.
    (b) End-to-end energy breakdown comparison across 3 representative PTC families under dynamic attention and static linear workloads with matched computations.
    All results are simulated using \textsc{SimPhony}~\cite{yin2024simphony}.
    \emph{Arch Setting}: Total of 8 8$\times$8 PTCs with 12 wavelengths, 8-bit precision, and a 5 GHz clock rate.
    \emph{Workload}: Attention uses sequence length $S=1024$ and embedding dimension $D=512$; Linear layer projects hidden dimension from 512 to 1024 with a batch of 4. 
    MZI mesh is incompatible with attention, thus omitted.
    }
    \label{fig:workload_study}

\end{figure}

\subsubsection{Simulation Results and Insights}
\label{subsubsec:sim_res}

Figure~\ref{fig:workload_study} presents the simulated end-to-end energy breakdown across the three representative PTC families, decomposed into key components: modulation (drivers), detection (readout), laser input, data conversion (ADC/DAC), and memory. Furthermore, it benchmarks these architectures against state-of-the-art (SOTA) electronic baselines, ranging from the NVIDIA A100 and H100 to the emerging B200.

These results highlight three critical insights:

\noindent
\textbf{(1) Photonic Competitiveness Against SOTA Electronics.}
The most distinct takeaway is the performance of the \textbf{TM-Crossbar} architecture. Explicitly designed to mitigate system taxes, specifically data movement and cross-domain conversion, this design demonstrates a competitive position on the density-efficiency Pareto frontier relative to the NVIDIA A100. Notably, it achieves superior energy efficiency even compared to the newer NVIDIA B200. This validates the trajectory shown in Fig.~\ref{fig:arch_efficiency}: when architectures are optimized to minimize peripheral overheads, photonics retains its fundamental advantage, suggesting even greater potential as designs move toward fully optical, constraint-free implementations.

\noindent
\textbf{(2) System Efficiency Limits Beyond the Optical Core.}~
In contrast, MZI meshes and MRR weight banks, which have been central to ONN research, can exhibit substantial \emph{system-level overheads} when evaluated end-to-end on modern workloads.
As illustrated by the breakdown in Fig.~\ref{fig:workload_study}(b), overall efficiency in these designs is often constrained by the surrounding system stack: high-speed ADC/DAC conversion, data movement, and programming/control overheads.
These costs become particularly prominent for dynamic and communication-heavy workloads (e.g., attention-style operations) where frequent reconfiguration and I/O amplify peripheral energy.

\noindent
\textbf{(3) The Rigidity of MZI Meshes in Dynamic Workloads.}
A critical limitation emerges regarding workload versatility: MZI meshes are fundamentally ill-suited for the Attention mechanisms. 
The MZI topology relies on a \emph{static} linear transform programmed via Singular Value Decomposition (SVD) and phase decomposition. 
In Attention, however, the effective operator is input-dependent and changes at every token step with no static weights. Since the SVD and phase decomposition cannot be precomputed and reprogramming an $N\times N$ mesh requires updating $\mathcal{O}(N^2)$ thermal or electro-optic phase shifters at token-rate timescales, MZI meshes are thermally and control-limited to static linear projections. Consequently, they lack the opportunity to scale to the dynamic workloads that define the current AI era.

\noindent
\textbf{Outlook.}~
It is important to note that the results in Fig.~\ref{fig:workload_study} represent "baseline" implementations of these topologies. The flourishing landscape of high-performance designs shown previously in Fig.~\ref{fig:arch_efficiency} is populated by works that explicitly target these identified bottlenecks, introducing novel optimizations in area, energy efficiency, and flexibility. In the following section, we categorize these recent advancements by mapping them to the specific scaling dimension they improve.

\section{Scaling Photonic AI: from Bottlenecks to Solutions}
\label{sec:photonic_ai}

\begin{figure}
    \centering
    \includegraphics[width=0.95\columnwidth]{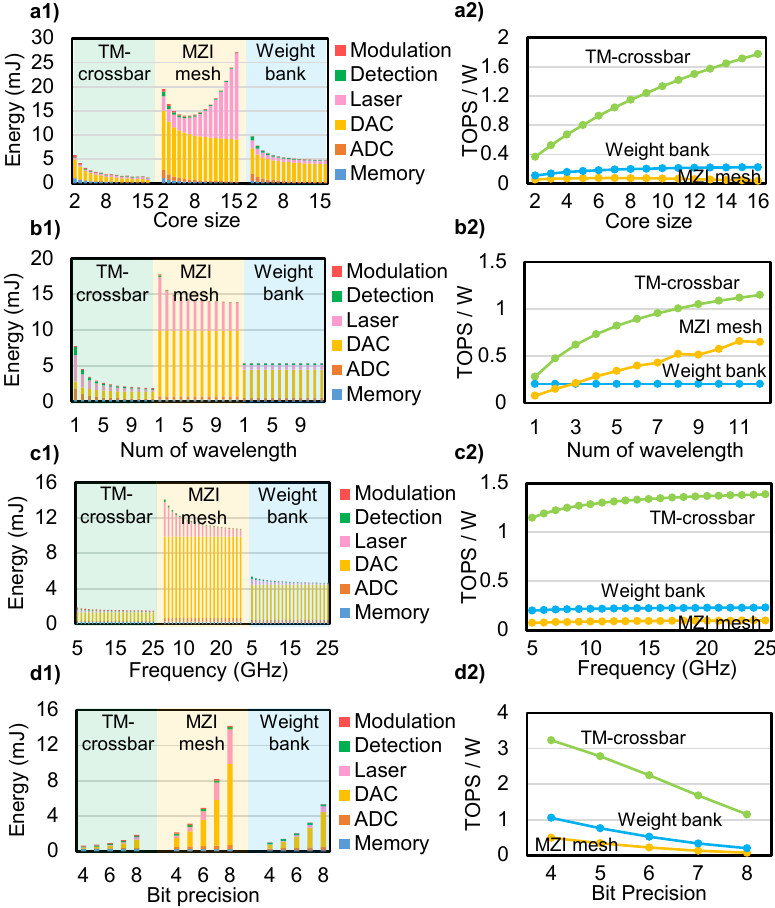}
    \caption{Impact of parameter scaling on energy breakdown and power efficiency across representative designs on \textbf{a single tensor core}. We evaluate three canonical photonic tensor core designs, the dynamic time-multiplexing crossbar~\cite{zhu2024lightening}, the Clements MZI mesh~\cite{shen2017deep}, and the weight bank~\cite{tait2016microring}, using simplified and standardized models. Energy breakdowns (left panels) and energy efficiency in TOPS/W (right panels) are reported as functions of (a) core size, (b) number of wavelengths, (c) bit precision, and (d) operating frequency. The baseline configuration is one 8$\times$8 core, 12 wavelengths, 8-bit precision, and 5 GHz operation. All results are obtained using SimPhony~\cite{yin2024simphony} to ensure a fair and consistent comparison across architectures. All scaling explorations are performed using a 4-batch linear projection ($512\!\rightarrow\!1024$) to ensure all PTC families can be evaluated.}
    \label{fig:param_scaling}

\end{figure}

\subsection{Identifying Bottlenecks via Key-Dimension Projections}
\label{subsec:param_projection}

In Sec.~\ref{subsubsec:sim_res}, we identified the ``system tax'' imposed by frequent data conversions and the rigidity of PTCs as primary barriers to efficient photonic computing. Extending this analysis, we utilize our simulation framework to sweep a broader set of architectural parameters, specifically \textit{operating frequency, bit precision, wavelength parallelism, and tensor core size}, to determine how to effectively fuse greater computational power onto a single chip and guide future scaling strategies.

We map these parameters to their physical impact on system performance. 
Tensor core size dictates the \textit{spatial density} of the compute fabric. 
Meanwhile, wavelength parallelism and operating frequency serve as complementary drivers of \textit{effective throughput}. Wavelength parallelism acts as a spectral multiplier, increasing the number of concurrent input channels supported on the same physical chip, which scales throughput analogously to increasing the temporal clock frequency. Critically, by sweeping bit precision, we assess the feasibility of supporting high-fidelity arithmetic within this mixed-signal paradigm.

Figure~\ref{fig:param_scaling} summarizes the simulated system-level energy breakdowns and energy efficiency trends across three PTC families, using the attention workload as a unified baseline. The results reveal distinct scaling behaviors across these dimensions:

\noindent\ding{202}~\textbf{Fusing Computational Density and Throughput:} As illustrated in Fig.~\ref{fig:param_scaling}(a) and (d), increasing the computational density, whether spatially via larger tensor cores or spectrally via dense wavelength division multiplexing (DWDM), generally yields improved energy efficiency. By increasing the number of wavelength channels, the system processes more data in parallel within the same footprint, effectively amortizing fixed static power overheads (such as laser biasing and thermal control) across a higher aggregate throughput.

\noindent\ding{203}~\textbf{Frequency Saturation:} While scaling the operating frequency (Fig.~\ref{fig:param_scaling}(b)) also improves throughput, the efficiency gains exhibit diminishing returns. 
At higher frequencies, the dynamic power consumption of high-speed drivers and readout circuits begins to scale (super-)linearly with the increased data rate, eventually neutralizing the efficiency benefits.

\noindent\ding{204}~\textbf{The Bit Precision Wall:} Most notably, Fig.~\ref{fig:param_scaling}(c) highlights a severe limitation: \textit{brute-force scaling of electronic bit precision is unsustainable}. As precision increases, energy consumption skyrockets while efficiency (TOPS/W) plummets. This is corroborated by the energy breakdown across all swept dimensions, where electro-optic conversion, dominated by DACs and optical modulation, emerges as the primary contributor to total energy consumption, consistently outweighing laser power and data movement. This confirms that the energy cost of high-resolution A/D conversion creates a fundamental ``precision wall'' for analog photonic computing.

\subsection{Categorizing Photonic AI Progress via Bottleneck-Driven Analysis} \label{subsec:progress_categorization}

\subsubsection{Area Efficiency: Device and Architectural Density}
Considering the relatively large footprint of optical components, device-level optimization can improve computing density by shrinking the overall area of EPICs. A device level approach is to use multi operand photonic primitives that collapse a length-$k$ dot product into a single physical unit, enabling accumulation directly in the physical domain concurrent with the electro-optic mapping $x_{\mathrm{out}}=f\!\left(\sum_{i=1}^{k} g(w_i, x_i)\right)$, where \(g(\cdot)\) captures operand encoding and \(f(\cdot)\) is the device transfer function. This strategy increases effective fan-in with a footprint comparable to single-operand optical synapses. Representative realizations include multi-operand MZIs, as well as MRRs and multimode inference devices (MMI)~\cite{feng2024integrated,gu2021squeezelight,ning2025microring,gu2024m3icro,li2026end}.

However, a key caveat is that multi-operand operation couples operands through the modulator nonlinearity, so individual contributions are less separable, which complicates calibration and training. The same nonlinearity can also be beneficial, since it can serve as an on-chip activation and reduce electrical post processing. Experiments with microring-based multi-operand neurons show that this nonlinear response can be trained and improve representational capacity, enabling comparable or better accuracy with fewer active modulators, which directly reduces both parameter count and tunable footprint~\cite{gu2021squeezelight,ning2025microring}.

Beyond device-level optimization, diffractive optical neural networks, DONNs, improve area efficiency at the architectural level by computing through diffraction and propagation across cascaded passive layers~\cite{wang2019chip}, optionally augmented with sparse active tuning. With modern nanofabrication, on chip DONNs based on metastructures or compact interferometric elements can implement wavefront shaping and support parallel Fourier and convolution like primitives in an ultra compact footprint~\cite{zhu2022space,meng2023compact,wang2022integrated}. Notably, this architecture achieves linear scalability in both footprint and energy consumption relative to input dimension, thereby circumventing the quadratic complexity overhead intrinsic to fully connected interferometric meshes. Recent hybrid diffractive–interference architectures further push efficiency while supporting larger-scale workloads~\cite{xu2024large,wang2025diffractive}.

However, a primary limitation of passive diffractive architectures is that their weight configuration is fixed upon fabrication, rendering them task-specific and generally incapable of realizing arbitrary matrix transformations on demand. Consequently, many architectures adopt hybrid strategies that combine diffractive propagation with programmable structures. In addition, partially reconfigurable diffractive designs have been demonstrated using post fabrication tuning elements such as microheater arrays~\cite{cheng2024multimodal,wang2025chip}. Nevertheless, realizing fully programmable, universal weight matrices in purely diffractive systems remains challenging, highlighting a trade-off between the compact efficiency of passive diffractive computing and the flexibility enabled by active photonic components.

\subsubsection{Throughput Scaling: Bandwidth and Multiplexing}
To optimize ONNs for high-throughput computing, a recurring strategy involves expanding parallelism by exploiting multiple orthogonal optical degrees of freedom, such as wavelength, time, and spacial multiplexing. By encoding data across these distinct dimensions, the architecture allows more MAC operations to be executed "in-flight" during a single propagation and detection cycle. An example is the universal optical vector convolution engine by Xu \emph{et al.}, which leverages an integrated microcomb source to concurrently harness temporal, wavelength, and spatial multiplexing, achieving $>$10~TOPS of computational throughput~\cite{xu202111}. More broadly, hyper-multiplexing architectures stack space–time–wavelength parallelism to reach trillions of operations per second while requiring only $\mathcal{O}(N)$ modulator devices to scale to $\mathcal{O}(N^2)$ operations per cycle~\cite{ou2025hypermultiplexed}. 
Related multiplexing-based approaches have also been reported~\cite{feldmann2021parallel,bai2025tops,xu2022high}. 
Along a complementary axis, Yin \emph{et al.} demonstrated an ONN that combines wavelength-division multiplexing (WDM) with mode-division multiplexing, using orthogonal spatial modes as an additional parallel channel to further boost on-chip throughput without proportionally increasing the device count~\cite{yin2023integrated}.

Complementary to exploiting orthogonal multiplexing modes, another throughput lever is the elevation of the electro-optic bandwidth ceiling, enabling time-serial streams to be processed at higher symbol rates. While typical integrated PTC report operation bandwidths in the GHz range, recent advancements in devices and material platforms allow for much more aggressive scaling. Lin \emph{et al.} demonstrated this potential by implementing a fully integrated tensor core based on thin-film lithium niobate (TFLN)~\cite{ou2025hypermultiplexed,lin2024120}, which supports in-situ weight updates at speeds over 40 GHz and a flexible fan-in.

\subsubsection{Energy Efficiency: Mitigating the Conversion Tax}
As noted above, in addition to area constraints, another major factor limiting ONN scalability is the energy overhead of electro-optic interfaces, as well as the practical challenges of calibration, control, and hardware complexity. While photonic platforms offer high-bandwidth processing, the power consumption required for driving modulators and performing data access and conversion (DAC/ADC) can dominate the total system energy budget, eroding the efficiency gains of optical computing. Consequently, a critical design objective is to minimize the information flux across the E-O boundary per inference, thereby restricting programmability to a sparse set of high-speed, low-power control elements. 

Returning to the context of diffractive ONNs, the diffractive backbone is typically passive and difficult to reconfigure at scale. Wang \emph{et al.} addressed this by placing a low-dimensional active modulation unit upstream of the passive diffractive cells~\cite{wang2025diffractive}. By injecting inputs through this compact active layer into the static diffractive volume, target transformations are synthesized via iterative time-domain updates. This approach strategically decouples the massive computational throughput from the control overhead, and shifts the programmability burden from a power-hungry, fully active spatial backbone to a minimal, rapid control layer. This significantly lowers the aggregate E-O interface cost while preserving the high-throughput advantages of diffractive propagation.

Another route to improving energy efficiency is to exploit the redundancy of modern DNNs to reduce the number of tunable parameters that must be physically encoded. Since high-accuracy solutions often reside in low-dimensional subspaces of the full weight space~\cite{han2015deep,molchanov2016pruning,ding2017circnn}, compressed parameterizations can directly translate into fewer active photonic degrees of freedom and fewer interface channels, which in turn lowers power on E-O modulation. Feng \emph{et al.} proposed the optical subspace neural network (OSNN), factorizing weights as $W=B\Sigma P$ with diagonal $\Sigma$ and implementing the unitary transforms $B$ and $P$ using hardware efficient butterfly meshes~\cite{feng2022compact}. By avoiding a fully programmable MZI mesh, the approach cuts the number of actively tuned elements by up to 7$\times$ while achieving 94.16\% accuracy on MNIST. Similarly, Ning \emph{et al.} imposed a block circulant constraint to realize structured compression~\cite{ning2025hardware}. Using a compact MRR based crossbar with WDM, the compression is embedded directly into the PIC topology, enabling up to a 75\% reduction in trainable parameters and control overhead with negligible accuracy degradation across multiple tasks. These approaches validate that strictly limiting the active E-O interface through structured, low-dimensional hardware design plays a pivotal role in achieving practical, energy-efficient photonic acceleration.

\noindent\textbf{Optical computing with non-volatile and analog memory.}~
To mitigate the prohibitive power consumption associated with signal conversion and state maintenance, recent research has explored augmenting photonic computing with non-volatile materials and analog memory structures.

On one hand, researchers aim to bypass the static power dissipation of traditional optical components, which rely on volatile mechanisms like electro-optic or thermo-optic effects to maintain their state. Instead, non-volatile devices, leveraging PCMs~\cite{feldmann2021parallel,zhang2023nonvolatile,xia2024seven}, ferroelectrics~\cite{geler2022ferroelectric}, or latching MEMS~\cite{errando2019mems, unamuno2005mems}, offer the ability to retain information without a continuous power supply.  However, existing non-volatile technologies, particularly PCMs, face endurance limitations depending on their modulation mechanism (electrical, electrothermal, or optical). This raises concerns regarding their suitability for write-intensive workloads like training~\cite{martin2022endurance}, although recent efforts are actively addressing these durability constraints~\cite{zhu2022elight}.

Complementary efforts focus on integrating foundry-compatible analog electronic memories, such as Dynamic Electro-Optic Analog Memory (DEOAM)~\cite{lam2026neuromorphicphotoniccomputingelectrooptic}, to resolve the interface bottleneck. In this approach, a capacitor is paired directly with an MRR to locally hold the drive voltage (data) on the modulator. By functioning as a "sample-and-hold" circuit, this architecture allows DACs to be time-multiplexed across columns of devices rather than requiring a dedicated DAC for every MRR. The system updates weights row-by-row while the analog memory retains the signal on the MRRs, thereby reducing the DAC count from quadratic ($N^2$) to linear ($N$) complexity and significantly relaxing energy and bandwidth constraints.

\subsubsection{Dynamic Workload Adaptation}

To elucidate the limitations of existing architectures, we first distinguish between \emph{static inference} and \emph{dynamic workloads}. Conventional inference relies on fixed weights, whereas dynamic workloads, most notably the attention mechanism in Transformers, require General Matrix Multiplication (GEMM) in which \emph{both operands vary at runtime}.

A canonical example is self attention:
\begin{equation}
\mathrm{Attn}(Q,K,V) = \mathrm{softmax}\left(\frac{QK^\top}{\sqrt{d_k}}\right)V,
\end{equation}
where the query ($Q$), key ($K$), and value ($V$) matrices are token dependent and generated on the fly. Unlike conventional layers, the $QK^\top$ operation induces dynamic, full range, all to all interactions between runtime generated operands.

More generally, we aim to support matrix multiplication of the form
\begin{equation}
\mathbf{Y} = \mathbf{W}\mathbf{X},
\end{equation}
where both $\mathbf{W}$ and $\mathbf{X}$ are dynamic, updated at runtime, and full range, containing both positive and negative values. These requirements impose stringent constraints that most legacy optical designs are fundamentally ill-equipped to satisfy.

\emph{Challenge 1: Reconfiguration Latency in Coherent MZI Meshes.}
Prior coherent architectures, such as Mach-Zehnder Interferometer (MZI) meshes~\cite{shen2017deep, Demirkiran2023}, struggle to efficiently support dynamic GEMM due to the high complexity of operand mapping. Unlike electronic crossbars, MZI meshes require unitary decompositions, most commonly SVD, to derive precise phase settings for each interferometric element. While acceptable for static weights, this process becomes a prohibitive runtime bottleneck when $\mathbf{W}$ changes every cycle. For example, computing the SVD and corresponding phase decomposition for a $12\times12$ matrix can take approximately 1.5~ms on a CPU, introducing system stalls that overwhelm the intrinsic speed of optical propagation. Moreover, to reduce footprint and insertion loss, these designs often rely on compact thermo-optic or non-volatile phase shifters, such as phase change materials. The programming latency of such devices, $10$~ns to $10$~$\mu$s, is orders of magnitude slower than the optical computation itself, rendering reconfiguration costs impossible to amortize under dynamic workloads.

\emph{Challenge 2: Sign Representation Overhead in Incoherent Architectures.}
Incoherent designs, such as MRR weight banks~\cite{tait2017neuromorphic}, face a fundamental limitation in representing full-range values. Because computation is performed through light intensity modulation, at least one operand must be non-negative. Supporting signed multiplication, therefore, requires decomposing operands into positive and negative components, for example, $X = X_+ - X_-$. A single multiplication $(X_+ - X_-)(W_+ - W_-)$ then expands into four sub operations, $X_+W_+$, $X_+W_-$, $X_-W_+$, and $X_-W_-$, which must be executed through time multiplexing or hardware duplication~\cite{NP_DAC2021_Sunny, NP_ISCA2021_shiflett}. This results in a $>2$ to $4\times$ increase in hardware complexity and energy consumption. By significantly increasing modulation, DAC, and control overheads, this decomposition erodes the efficiency benefits that incoherent architectures typically derive from weight static dataflows.

\noindent\textbf{Existing solutions for dynamic workloads.}~
With the emergence of Transformer-based foundation models, the dominant computational paradigm has shifted from static convolutions to dynamic attention mechanisms. In this regime, hardware must efficiently support interactions between runtime-generated operands, rather than between fixed weights and activations.

To address this challenge, Zhu \emph{et al.} proposed \emph{Lightening Transformer}~\cite{zhu2024lightening}, a high-speed optical accelerator that co-designs the photonic datapath with a specialized on-chip and off-chip tiling strategy. The system accelerates dynamic attention by leveraging coherent interference and WDM to achieve high spectral parallelism. To alleviate data movement bottlenecks, the architecture adopts a crossbar-style topology that maximizes intra-core operand reuse, employing dynamic orchestration of broadcast and tiling to amortize electro-optical conversion costs.

Complementary efforts have explored time domain integration to support dynamic functionality. Rahimi \emph{et al.} proposed a time multiplexed realization for executing dynamic dot products~\cite{rahimi2024realization}, sharing architectural principles with the optical tensor processor introduced by Yin \emph{et al.}~\cite{yin_dac2025}.

Beyond attention accelerators, recent work from Lightmatter demonstrated a universal photonic processor capable of supporting a wide range of workloads~\cite{ahmed2025universal}, including natural language processing and deep reinforcement learning. To overcome the intrinsic rigidity of optical weight encoding, their design shifts weight programmability to the electrical domain using a differential photodetection unit coupled with a resistive, differential DAC. This approach enables full reconfigurability without incurring the latency penalties associated with tuning thermal or phase change optical elements.

\section{Enabling Photonic AI at Scale: Full-Lifecycle Electronic-Photonic Design Automation}
\label{sec:epda}
Photonic artificial intelligence systems demand a level of design complexity and scale that can no longer be supported by conventional ad hoc, isolated, manual design flow.
As photonics moves toward very-large photonic integration (VLPI) and heterogeneous EPICs, successful deployment increasingly depends on a full-lifecycle EPDA stack.
We focus on how advances in co-simulation, inverse and automated design, and physical design automation are converging to enable photonic AI systems that are not only high-performing, but also manufacturable, robust, and deployable at scale.

\subsection{EPDA: Device-/Circuit-Level Capabilities}
In an EPDA stack, device- and circuit-level simulation is the \emph{translation layer} between nanophotonic physics and system-level performance.
For photonic machine intelligence, simulation is not merely verification, but also enables co-design across \emph{devices, circuits, architectures, and learning algorithms}.
Thus, ``good simulation'' must provide \emph{scalability} to large design, \emph{composability} into system, \emph{parametric conditioning} over operating conditions, and \emph{differentiability/uncertainty awareness} for inverse design and robustness, motivating AI-assisted alternatives.

\subsubsection{AI-assisted photonic device simulation}
AI-assisted simulation uses ML to approximate the Maxwell solution operator, addressing the oversimplification of compact models and the scalability limits of rigorous electro-magnetic (EM) solvers.
These surrogates seek to preserve physical accuracy while delivering the speed, scalability, and differentiability needed for EPDA, especially for \emph{system-aware} workflows such as parametric sweeps, composable circuit co-simulation, and differentiable inverse design.
This EM surrogate maps a device description, typically discretized permittivity plus sources and boundary conditions, to EM responses.
Table~\ref{tab:AI_simulation} summarizes representative approaches by domain (frequency vs.\ time), physics priors, and scalability.
\begin{table*}[]
\caption{Categorized AI-assisted photonic device full-field simulation approaches with key feature comparison. \emph{N-} means normalized. MAE is the mean absolute error.}
\label{tab:AI_simulation}
\resizebox{\textwidth}{!}{
\begin{tabular}{c|c|ccccccc}
\hline
 Domain                                                                       & Approach                  & Model                                                                                            & Inputs | Condition            & Physics Prior         & Devices/Scales                                   & Speedup                                              & Error          & Comment     \\ \hline
\multirow{5}{*}{\begin{tabular}[c]{@{}c@{}}Frequency\\ -domain\end{tabular}} & \multirow{4}{*}{One-shot} & MaxwellNet~\cite{NP_APL2022_Lim}                                                                                       & $\epsilon_r~|~\lambda,\Omega,J,$ PML   & Maxwell Loss          & Lens/$<10\lambda$                                             &  300-600$\times$ over COMSOL                                                 &  $\sim$0.01 N-L2 Error              &                 \\
                                                                                                                                                          &                           & WaveY-Net~\cite{NP_Nature2021_Chen} (U-Net)                                                                                 & $\epsilon_r~|~\lambda,\Omega,J,$ PML   & Maxwell Loss          & Grating/$<10\lambda$                                          & 700$\times$ over Direct Solver               & 3e-2 MAE       &  \multirow{4}{*}[1.5em]{\begin{tabular}[c]{@{}c@{}}\textcolor{uclablue}{Fast}\\ \textcolor{mydarkred}{Limited scalability}\end{tabular}}               \\
                                                                                                                                                          &                           & NeurOLight~\cite{gu2022NeurOLight} (FNO)                                                             & $\epsilon_r,\lambda,\Omega,J~|$ PML & Wave Prior            & Tunable/etched MMI/$\sim 10 \lambda$           & \textgreater{}100$\times$ over Direct Solver & 0.12 N-MAE     &                 \\
                                                                                                                                                          &                           & PACE~\cite{zhu2024pace} (FNO)                                                                   & $\epsilon_r, \lambda,\Omega,J~|$ PML & Wave Prior            & Tunable/etched MMI, Metaline/$\sim 10 \lambda$ & 150-500$\times$ over Direct Solver           & 0.03-0.1 N-MAE &                 \\ \cline{2-9} 
                                                                                                                                                        & Iterative  & FNO+F-GMRES~\cite{mao2025accuratescalabledeepmaxwell} & $\epsilon_r, \lambda,\Omega,J, |Ae-b|,$ PML      & Maxwell Residual      & WDM, Coupler, Metalens/$\sim100\lambda$           & 10$\times$ over iterative GMRES                    & 1e-3 L1 Error  & \begin{tabular}[c]{@{}c@{}}\textcolor{uclablue}{Scalable}\\ \textcolor{mydarkred}{Limited speedup}\end{tabular} \\ \hline 
                                                                            \begin{tabular}[c]{@{}c@{}}Time\\ -domain\end{tabular}                       &       Autoregressive                    & PIC$^2$OSim~\cite{ma2025pic2o} (CNN)                                                                           & $\epsilon_r,\lambda,\Omega,J$~|~PML          & Causality, Model & MRR, MMI, Metaline                               & 300-600$\times$ over MEEP                          & 3e-2 N-L2Norm  & \begin{tabular}[c]{@{}c@{}}\textcolor{uclablue}{Broadband}\\ \textcolor{mydarkred}{Error accumulates}\end{tabular}                 \\ \hline
\end{tabular}
}
\end{table*}

\noindent\textbf{Prediction target: FoM vs. EM field.}~
\uline{FoM surrogates} predict metrics (e.g., S-parameters, group index) without full-field reconstruction.
They are fast but often \emph{non-composable} and device-specific.
\uline{EM field surrogates} predict steady-state fields or time-domain evolution, enabling downstream multiple FoMs extraction and inverse design; their \textbf{composability}, \textbf{generalization}, and \textbf{differentiability} are essential to circuit- and system-level co-simulation.

\noindent\textbf{Frequency-domain field surrogates: one-shot vs. iterative.}~
Most methods target steady-state \emph{frequency-domain} solutions.
\uline{One-shot} models map devices/conditions to complex fields in one pass; \emph{physics-driven/augmented} CNNs (e.g., MaxwellNet~\cite{NP_APL2022_Lim}, WaveY-Net~\cite{NP_Nature2021_Chen}) use residuals (e.g., $|Ae-b|$) to reduce artifacts while accelerating inference.
A second line of works, \uline{Operator learning}~\cite{Azizzadenesheli2024NeuralOperators}, approximates \textbf{parametric} Maxwell operators. 
NeurOLight~\cite{gu2022NeurOLight} enables fast sweeps over wavelength, sources, and permittivity. PACE~\cite{zhu2024pace}) improves fidelity on challenging structures such as metalines and large interferometric devices.
From an EPDA perspective, one-shot surrogates can reach $10^2\sim10^3\times$ speedups but often degrade on larger domains or complex scattering~\cite{NP_APL2022_Lim}; transfer learning~\cite{gu2022NeurOLight} helps, yet robust generalization remains open.

To address domain scaling,
\uline{Iterative Maxwell neural solving} hybridizes learned local solves with classical loops (often via domain decomposition) and refines until criteria such as $|\mathbf{A}x-b|<\epsilon$, trading smaller speedups (often $\sim 10\times$) for robustness on large ($>100\lambda$) domains.

\noindent\textbf{Time-domain field surrogates.}~
Time-domain surrogates learn FDTD-like spatiotemporal evolution~\cite{Zhang2025TimeDomainEM, ma2025pic2o} for transient/broadband behavior, but must address \emph{long-horizon} error accumulation.
PIC$^2$O-Sim~\cite{ma2025pic2o} uses causality-aware dynamic convolution aligned with Maxwell dynamics to achieve large speedups over FDTD (e.g., MEEP) with stable rollouts.

\noindent\textbf{Physics priors and learning paradigms.}~
Across the aforementioned methods, the central question is \uline{how physics is incorporated}.
\emph{Physics-driven} residual minimization reduces labels but can be optimization unstable. 
\emph{Physics-augmented} training adds PDE/boundary and conservation/reciprocity constraints to improve physical validity.
\emph{Data-driven} operator learning (NeurOLight~\cite{gu2022NeurOLight}, PACE~\cite{zhu2024pace}, PIC$^2$O-Sim~\cite{ma2025pic2o}) embeds physics via \textbf{inputs}, \textbf{architectures} (local causality, global interference), and \textbf{training} (e.g., superposition augmentation~\cite{gu2022NeurOLight}) for better generalization.

\subsubsection{Photonic-electronic circuit-level co-simulation}
After device-level validation, circuit-level simulation captures component interactions for system modeling.
A co-simulator must balance (i) enough fidelity to model nonidealities (loss/dispersion, reflections/feedback, modulation limits, noise, thermal drift) and (ii) enough speed for architecture exploration and learning-hardware co-design.

\noindent\textbf{The prevailing workflow: extract-then-simulate.}~
Today’s dominant flow is \emph{hierarchical}: devices are characterized by EM/measurement, reduced to compact models (e.g., frequency-dependent S-parameters), and composed in circuit simulators for sweeps and link budgets.
Electronics (drivers, TIAs, control, DAC/ADC) are usually simulated separately in SPICE/Verilog(-A) with coarse interfaces.
This \textbf{\emph{split-flow} breaks down} for large photonic AI systems where mixed-domain interactions (loading, quantization/noise, feedback/calibration, thermal drift) jointly set performance, highlighting the need for unified, scalable photonic-electronic co-simulation~\cite{SPIPE}.

\noindent\textbf{Challenges in EPIC co-simulation.}~
The difficulty is rooted in a \emph{mismatch of native formalisms}: electronics solvers operate on voltages/currents in time-domain modified nodal analysis, while photonic circuits use complex waves and multiport scattering across wavelength/polarization.
Bridging them needs stable, physically consistent interface models (modulators, detectors, impedance/parasitics) that translate electrical-optical variables~\cite{SPIPE}, which is why many flows either couple domains manually or translate one into the other.

\noindent\ding{202}~\textbf{Circuit-level Unification via Behavioral Compact Models.}~
Compact models make large-scale PIC simulation tractable and act as a \emph{process development kit (PDK) contract}, typically via S-parameters for fast composition and sweeps.
A common strategy implements photonics as behavioral models in electronics-native languages (Verilog-A~\cite{oe2015_veriloga_cosim, NP_MWSCAS2019_Shawon}, SPICE~\cite{EPICSPICE}) so photonic blocks run inside mature electronic design automation (EDA) simulators, improving interface realism and mixed-signal verification~\cite{NP_MWSCAS2019_Shawon}.
This enables unified transient analysis and device-specific models for co-design with CMOS drivers/receivers, but introduces costs: model translation can be labor-intensive and inconsistent with PDK models, and broadband response can require expensive sweeps (mitigated by chirp-based transients)~\cite{NP_MWSCAS2019_Shawon}.
Thus, compact-model unification is necessary but not sufficient for faithful mixed-domain validation at photonic-AI scale.

\noindent\ding{203}~\textbf{Coupled-Domain Co-simulation that Preserves Native Abstractions.}~
An alternative couples domain-appropriate solvers through explicit interfaces instead of forcing a single representation, avoiding staged ``hand-off'' co-simulation.
This becomes crucial as systems exhibit \emph{nonlinear, time-variant} electro-photonic interactions and feedback~\cite{oe2015_veriloga_cosim}.
Recent work uses microwave-style \emph{power waves} for bidirectional/reflection-aware modeling~\cite{OE2024_BidirectionalVA}, and SPIPE couples a SPICE engine with an S-parameter photonic solver via physical modulator/photodetector interfaces~\cite{SPIPE}, preserving transistor-level transients and scalable photonic S-matrix composition.

\noindent\textbf{Perspectives and Future Directions.}~
Looking forward, circuit-level EPIC simulation must move beyond ``simulate a netlist'' toward \emph{simulate and optimize heterogeneous EPIC systems under realistic conditions}.
As photonic AI hardware scales, key directions include:
\textbf{(1) Scalability}: co-simulation over thousands of devices with complex coupling (feedback, monitoring, reconfiguration);
\textbf{(2) Radio frequency (RF)- and multi-wavelength awareness}: jointly modeling bandwidth/impedance/RF-optical effects with wavelength-dependent propagation and interference;
\textbf{(3) Differentiability}: enabling end-to-end gradients for joint device/circuit optimization under AI-centric metrics;
\textbf{(4) Layout awareness}: post-layout back-annotation via extracted parasitics and interconnect models for layout-dependent effects; and
\textbf{(5) Measurement-in-the-loop}: measurement-informed digital twins that update compact models and uncertainty bounds to improve variation robustness.

\subsection{EPDA: Architecture-Level Modeling}

While device- and circuit-level simulation captures the physics of individual photonic components, it is insufficient for \emph{system-valid} evaluation of photonic AI accelerators, where end-to-end efficiency emerges from the interaction among photonic compute units, electronic peripherals, conversion interfaces, memory hierarchy, interconnects, calibration/control, and workload mapping. 
Architecture-level EPDA provides the abstraction layer that connects component characteristics to system metrics under realistic workloads. 
Crucially, for photonic machine intelligence, architecture modeling must go beyond conventional ``performance modeling'' and explicitly represent optics-specific parallelism, analog error sources, and cross-domain overheads that can dominate at scale.

\subsubsection{What makes photonic architecture modeling different?}
\label{sec:epda_arch_motivation}
Architecture-level EPDA aims to model latency, throughput, energy, area, and accuracy trade-offs of heterogeneous EPIC systems without resolving electromagnetic fields. However, photonic AI systems violate several implicit assumptions that underpin many electronic-accelerator simulators:
\begin{itemize}
    \item \textbf{Cross-domain overheads are first-order.} Electrical-optical (E-O) interfaces (e.g., DAC/ADC, modulators, TIAs) can outweigh optical compute energy when scaled to high bandwidths or high precision.
    \item \textbf{Optics introduces non-digital error behavior.} Phase noise, laser relative intensity noise, drift, interference, and analog accumulation yield \emph{correlated} and often \emph{data-dependent} errors, which cannot be captured by simple bit-flip or i.i.d.\ additive-noise models.
    \item \textbf{Parallelism is multi-dimensional.} Spatial replication, WDM, and broadcast/accumulate structures alter utilization, scheduling, and bottlenecks in ways that do not map cleanly to standard systolic or SRAM-centric models.
\end{itemize}
Therefore, a credible EPDA stack must (i) account for conversion and control costs, (ii) connect non-idealities to algorithm-level accuracy, and (iii) model workload-to-hardware mapping with photonics-aware primitives.

\subsubsection{Adapting electronic architecture simulators}
Early efforts toward architecture-level photonic modeling leveraged structural similarities between photonic accelerators and analog compute-in-memory (CiM) systems. 
By adapting existing CiM architectural simulators, researchers demonstrated that photonic systems can be evaluated within a full-system context that accounts for DRAM access, on-chip buffering, and cross-domain data movement. 
This line of work provided an important system-level insight: even when optical-domain computation is highly efficient, data conversion and memory traffic can dominate total system energy, highlighting the need for joint consideration of architecture and mapping strategies.

These CiM-inspired approaches primarily emphasize array-style accelerator organizations and dataflow-centric analysis, which are inherited from electronic architectures. 
While effective for capturing full-system behavior and enabling rapid design-space exploration, they are generally tailored to regular compute structures and abstract photonic hardware at a coarse architectural level. 
As a result, they are particularly well-suited for system-level comparisons and workload-driven analysis, rather than detailed exploration of photonic-specific architectural diversity.

\subsubsection{Architecture-specific photonic simulators}
Beyond generic CiM-based modeling, several photonic accelerator efforts have developed custom architecture-level simulators tailored to specific optical computing paradigms. 
A representative example is the coherent photonic crossbar accelerator based on PCM, which employs a modified SCALE-Sim–based framework to model compute cycles, weight programming overhead, memory accesses, and peripheral electronics at the system level.

In this approach, cycle-accurate architectural modeling is combined with simulated/measured device characteristics, including losses, laser efficiency, ADC/DAC power, and SRAM/DRAM access energy, to evaluate end-to-end metrics such as throughput, energy efficiency, and chip area for large convolutional neural network workloads.

By explicitly modeling programming latency, batch size, and memory hierarchy effects, this class of simulators demonstrates how system-level constraints significantly influence the scalability of photonic accelerators beyond small arrays.

At the same time, these architecture-specific simulators are intentionally optimized around a given photonic design and operating regime. 
This specialization enables detailed and realistic evaluation of targeted architectures, while naturally limiting direct reuse for exploring a wide range of PTC topologies or performing broad cross-architecture comparisons.

\subsubsection{Native photonic architecture modeling frameworks}
More recently, photonic architecture modeling frameworks have emerged that aim to unify device behavior, circuit organization, and architectural execution within a unified flow (e.g., cross-layer approaches such as SimPhony~\cite{yin2024simphony}). 
Instead of assuming a fixed array abstraction like in electronic accelerators, these frameworks enable \emph{parametric} construction of heterogeneous PTCs, encompassing mesh-, array-, and broadcast-style structures under a unified representation. 
At the architectural level, they explicitly model \emph{photonic dataflow}, including multi-dimensional parallelism along spatial, spectral, and temporal dimensions, and hierarchical mixed-signal accumulation patterns. 
In addition, energy and performance estimation is often tied to configuration states, workload characteristics, and hardware budgets, allowing loss, laser power, and footprint to be reflected directly in architectural evaluation. 
By making such assumptions explicit, photonics-native frameworks facilitate more systematic and transparent architectural exploration across a wider design space.

Nevertheless, like architecture-specific simulators, these frameworks inevitably rely on behavioral abstractions and modeling assumptions whose validity depends on how non-idealities, calibration procedures, and control overheads are represented. Effects such as thermal crosstalk, fabrication variation, wavelength drift, and coherence-related accuracy degradation are typically incorporated only approximately.

\subsubsection{Open problem and future direction}
Architecture-level EPDA for photonic AI is still at an early stage: most existing studies provide valuable proof-of-concept system modeling and design space exploration~\cite{NP_ICCAD2021_Li_OHAS}, yet the field lacks a \emph{full-lifecycle} methodology that connects workload intent to implementable heterogeneous EPIC systems with predictable performance and closed-loop co-design.
Looking forward, the key opportunity is to elevate architecture-level EPDA from ``estimating throughput on ideal blocks'' to \emph{application-to-hardware co-design and compile-time system synthesis}, grounded in physically realistic constraints and validated through cross-layer closure.

\noindent\ding{202}~\textbf{Application-architecture co-design.}~
Future photonic accelerators will be judged by end-to-end workload performance.
This requires EPDA frameworks that understand the structure of modern applications, AI, scientific computing, and beyond, and expose architectural knobs that matter in practice: tensor operator support, precision formats, activation/dataflow movement, and control/calibration scheduling.
A central research direction is to build \emph{workload-aware photonic architectures}, where the choice of photonic compute primitive is guided by application structure.

\noindent\ding{203}~\textbf{Workload-to-system compiling and mapping for heterogeneous EPIC.}~
A missing layer in many photonic architecture studies is a compiler-grade mapping stack that translates models into executable schedules under heterogeneous constraints:
photonic resource allocation, memory hierarchy and data movement, conversion and control overheads, and timing constraints for reconfiguration.
Prior work \texttt{H$^3$PIMap} leverages \textsc{SimPhony}as a performance evaluator to optimize workload mapping for 3-D hybrid photonic in-memory computing systems~\cite{yin2026h3pimapheterogeneityawaremultiobjectivednn}.
In the future, EPDA needs \emph{workload-to-system compilation} that co-optimizes dataflow, tiling, placement of compute/memory, and reconfiguration policies.

\noindent\ding{204}~\textbf{Rigorous system performance evaluation: from ideal operations to signal integrity and robustness.}~
Architecture-level evaluation must move beyond idealized TOPS/W estimates and simplistic independent noise injection.
Photonic systems are constrained by signal integrity (SNR, effective number of bits (ENOB), dynamic range, bandwidth), polarization/wavelength/thermal management.
A major open problem is to develop \emph{architecture-appropriate abstractions} that capture these effects with high fidelity and efficiency and to support uncertainty-aware evaluation.

In summary, the next phase of architecture-level EPDA will be defined by \emph{compiler-like workload mapping}, \emph{physically grounded system evaluation}, and \emph{closed-loop cross-layer co-design} that links architectural intent to implementable and robust heterogeneous photonic-electronic systems.

\subsection{EPDA: Component-level Inverse Design}
\begin{figure}
    \centering
    \includegraphics[width=\columnwidth]{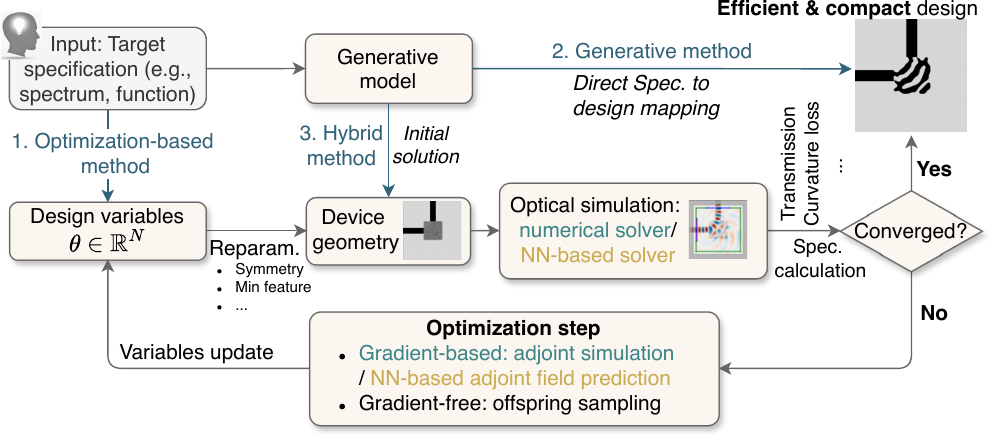}
    \caption{Photonic component inverse design flows. (1) Optimization-based, (2) Generative, and (3) Hybrid methods.}
    \label{fig:IDflow}
\end{figure}

\begin{table*}[]
\centering
\caption{Representative inverse-designed photonic devices, comparing geometry parameterization schemes, optimization methods, degrees of freedom (DoF), design-region sizes, fabrication-awareness strategies, and the approximate number of simulations required (\# Sims). 
Devices marked with $^\dagger$ are active components. \textit{Abbreviations:} PSO = particle swarm optimization; BO = Bayesian optimization; GA = genetic algorithm; DRL = deep reinforcement learning; DBS = direct-binary search; AGO = adjoint gradient optimization.}
\resizebox{0.9\textwidth}{!}{
\begin{tabular}{|c|c|c|c|c|c|c|}
\hline
\textbf{Device}                          & \textbf{Geometry} & \textbf{InvDes method} & \textbf{DoF} & \textbf{Design region}               & \textbf{Fab-aware}      & \textbf{\# Sims} \\ \hline
Optical amplifier$^{\dagger}$~\cite{zhao_highly_2023} & Structural                       & PSO + NN              & 7            & N/A                                   & \xmark                            & $\sim$10,000      \\
Wavelength router~\cite{wang_efficient_2024}                        & Structural                       & PSO + NN              & 5            & N/A                                   & Post-hoc                        & $\sim$1,000       \\
Phase shifter~\cite{liao_inverse_2023}                            & Structural                       & PSO                   & 3            & Lenght $\sim 3~\mu\mathrm{m}$         & \xmark                            & $\sim$1,000       \\
Few-mode fiber~\cite{chebaane_machine_2024}                           & Structural                       & Inversed NN           & 5            & N/A                                   & Post-hoc                        & $\sim$10,000      \\ \hline
Microring   resonator~\cite{gao2022BO}                    & Boundary                         & BO                    & $\sim$10     & Radius $\sim 3~\mu\mathrm{m}$         & Post-hoc                        & $\sim$100        \\ 
Mode splitter~\cite{liao_inverse_2024}                            & Boundary                         & AGO                   & $\sim$200    & $ 14\times2.5~\mu\mathrm{m}^2$        & Post-hoc                        & $\sim$400       \\
Power splitter~\cite{liu_inverse_2025}                           & Boundary                         & AGO                   & $\sim$200    & $6\times2.7~\mu\mathrm{m}^2$          & $\Delta y$ \text{bound}                    & $\sim$200        \\ \hline
Integrated lens~\cite{marques-hueso_genetic_2010}                          & Element-array                    & GA                    & $\sim$100    & $\sim5\times10~\mu\mathrm{m}^2$       & \xmark                            & $\sim$1,000       \\
Nanobeam laser~\cite{li_deep_2023}               & Element-array                    & DRL                   & $\sim$200    & $20\times0.7~\mu\mathrm{m}^2$         & \xmark                            & $\sim$1,000       \\
Grating coupler~\cite{sapra_inverse_2019}                          & Element-array                    & AGO                   & $\sim$200    & $12\times0.5~\mu\mathrm{m}^2$         & Feature-size      & $\sim$600        \\ \hline
Silicon modulator$^{\dagger}$~\cite{zhu_pso-aided_2024} & Pixel-based                      & PSO                   & $\sim$30     & N/A                                   & \xmark                            & $\sim$1,500       \\
Polarization rotator~\cite{yu_genetic-algorithm-optimized_2017}                     & Pixel-based                      & GA                    & $\sim$280    & Length $\sim 3~\mu\mathrm{m}$         & \xmark                            & $\sim$10,000      \\
Four-mode crossing~\cite{muratsubaki_direct-binary-search_2022}                       & Pixel-based                      & DBS                   & $\sim$2,000   & $\sim 12 \times 12~\mu\mathrm{m}^2$   & Hole diameter           & $\sim$1,000       \\
MVM unit~\cite{wang_inverse-designed_2024}                                 & Pixel-based                      & AGO                   & $\sim$1,000   & $4.8\times2.88~\mu\mathrm{m}^2$       & Pattern averaging         & $\sim$400       \\
PCM MMI$^{\dagger}$~\cite{wu_reconfigurable_2025}          & Pixel-based                      & AGO                   & $\sim$1,000   & $\sim40\times8.5~\mu\mathrm{m}^2$     & \xmark                            & $\sim$500       \\
Power splitter~\cite{Tang2020GenerativeNanophotonics}                           & Pixel-based                      & Generative NN         & $\sim$400    & $2.25\times2.25~\mu\mathrm{m}^2$      & \xmark                            & $\sim$10,000      \\ 
Wavelength filters~\cite{mao2023multi}                       & Pixel-based                      & AGO + NN              & $\sim$500    & $4\times2~\mu\mathrm{m}^2$            & Feature-size             & $\sim$20,000      \\  \hline
WDM demultiplexer~\cite{piggott_inverse_2015}                        & Free-form                        & AGO                   & $\sim$10,000  & $2.8\times2.8~\mu\mathrm{m}^2$        & Multi-$\lambda$ broadband             & $\sim$400       \\
MVM unit~\cite{nikkhah_inverse-designed_2024}                                 & Free-form                        & AGO                   & $\sim$10000  & $\sim11\times10.3~\mu\mathrm{m}^2$    & Low-index contrast              & $\sim$500       \\
Nonlinear optical switch~\cite{hughes_adjoint_2018}                 & Free-form                        & AGO                   & $\sim$20,000  & $\sim5\times5~\mu\mathrm{m}^2$        & Feature-size             & $\sim$2,000             \\
WDM demultiplexer~\cite{mao2023multi}                        & Free-form                        & AGO                   & $\sim$10,000  & $6.4\times6.4~\mu\mathrm{m}^2$        & In-loop DRC                     & $\sim$1,000    \\ \hline
\end{tabular}
}
\label{tab:inv}
\end{table*}

\begin{figure}
    \centering
    \includegraphics[width=\columnwidth]{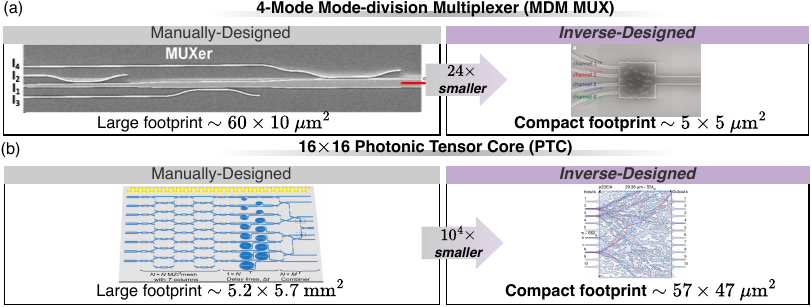}
    \caption{Inverse-designed photonic components and circuit modules can achieve similar functionalities with orders-of-magnitude smaller spatial footprint compared to manual counterparts. 
    (a) Manual~\cite{wang2013silicon} and inverse-designed four-mode mode-division multiplexer~\cite{yang2022multi}. 
    (b) Manual~\cite{xie2025complex} and inverse-designed photonic tensor core circuit~\cite{nikkhah_inverse-designed_2024}.
    }
    \label{fig:inverse_mot}
\end{figure}

\begin{figure}
    \centering
    \includegraphics[width=\columnwidth]{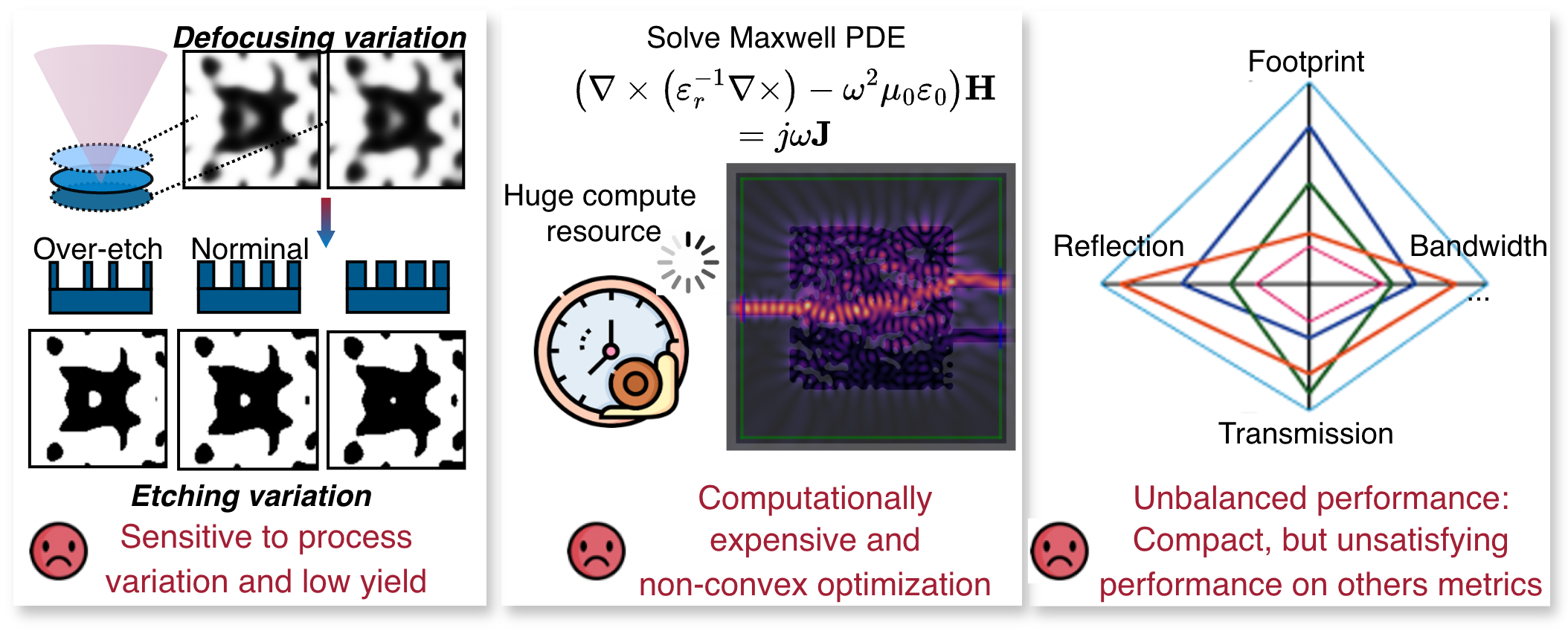}
    \caption{Challenges of photonic inverse design. }
    \label{fig:IDchallenges}
\end{figure}

\subsubsection{Limitations of manually designed devices}
Conventional photonic device design largely follows a \emph{forward-design} workflow: starting from canonical topologies (e.g., couplers, rings) and tuning a small set of geometric parameters via simulation sweeps.
While effective for standard building blocks, it becomes a bottleneck when photonic AI systems require \emph{compact footprints}, \emph{complex functionality}, and \emph{multi-metric optimization} (loss, bandwidth, extinction).
Manual tuning typically searches only a \uline{\emph{low-dimensional subspace of a chosen topology}}, limiting discovery of non-intuitive structures and often forcing larger area or degraded performance under tight constraints.
It is also \uline{\emph{expert-dependent}}, relying on significant intuition and trial-and-error that hinders accessibility and scalability.

These limitations diverge from the trend in electronics, where automation has turned design into a computation-driven workflow that scales with complexity~\cite{wang2009electronic, huang2021machine}.
Photonics has not yet fully leveraged modern compute and AI/EDA advances, motivating \emph{automated device synthesis} as an optimization problem that expands the search space and directly targets circuit- and system-level objectives.
    
\subsubsection{Introduction to inverse design of photonic devices}
To overcome the limits of manual trial-and-error, \textbf{inverse design} starts from a \emph{target specification} (e.g., spectrum or figure of merit (FoM)) and automatically synthesizes device geometry within a design region (Fig.~\ref{fig:IDflow}).
A typical pipeline chooses design variables $\boldsymbol{\theta}\in\mathbb{R}^{N}$ and a parameterization mapping $\boldsymbol{\theta}$ to layout (pixels, splines, etc.), evaluates the FoM via a Maxwell solver or surrogate, and iteratively updates $\boldsymbol{\theta}$ using adjoint gradients or gradient-free search until convergence.
By exploring high-dimensional, non-intuitive spaces, inverse design can produce compact structures that are hard to obtain by hand, often achieving similar functionality with a much smaller footprint (Fig.~\ref{fig:inverse_mot}), enabling dense photonic integration.

\noindent\underline{\textbf{Challenges in photonic inverse design.}}~
Although photonic inverse design has demonstrated strong potential,
key challenges remain (Fig.~\ref{fig:IDchallenges}):
\ding{202}\textbf{Manufacturability/yield:} irregular or pixelated layouts can violate foundry rules and increase process sensitivity (e.g., etch bias, line edge roughness, linewidth variation), thereby degrading yield;
\ding{203}\textbf{Simulation cost:} repeated high-fidelity solves across wavelengths/polarizations/corners are computationally expensive;
\ding{204}\textbf{Non-uniqueness/non-convexity:} many-to-one mappings, where distinct geometries can produce similar responses, and local minima make outcomes initialization-dependent; and
\ding{205}\textbf{Multi-objective trade-offs:} improving one metric can hurt other metrics (e.g., bandwidth, loss, or crosstalk), requiring principled objective balancing, constraint- and application-aware optimization.

\noindent\underline{\textbf{Categorization of photonic inverse design methods.}}~
To address the challenges, inverse-design methods broadly fall into \textbf{(i) optimization-driven} and \textbf{(ii) AI-assisted} families~\cite{Su2025MLPhotonicsEDA,2025_Marzban_Invdes}, which are often combined to balance exploration and efficiency.

\emph{Optimization-driven approaches} typically include \textbf{heuristic/evolutionary} and \textbf{gradient-based} methods.
Heuristics offer \textbf{global exploration} by maintaining a population of candidates (GA~\cite{xie2020design, marques-hueso_genetic_2010, yu_genetic-algorithm-optimized_2017}, PSO~\cite{zhang2023improved, zhao_highly_2023, wang_efficient_2024, liao_inverse_2023, zhu_pso-aided_2024}, DBS~\cite{hansi2025silicon, muratsubaki_direct-binary-search_2022}), well suited to low-dimensional or discrete designs, but are often simulation-hungry (e.g., $\sim$100 hours for GA in some cases~\cite{xie2020design}); Bayesian Optimization (BO)~\cite{gao2022BO} improves sample efficiency but struggles as dimension grows.

\emph{Gradient-based inverse design} typically uses adjoint method~\cite{givoli2021tutorial} to obtain $\partial \mathrm{FoM}/\partial\boldsymbol{\theta}$ for thousands of parameters with one extra simulation, enabling large-scale topology optimization~\cite{wang_inverse-designed_2024, wu_reconfigurable_2025, mao2023multi, piggott_inverse_2015, nikkhah_inverse-designed_2024, hughes_adjoint_2018, schubert_inverse_2022}; however, it is \textbf{local and initialization-sensitive}, and can yield non-manufacturable or variation-fragile patterns without constraints/regularization~\cite{Ma2025BOSON}.

\emph{AI-based methods} typically include \textbf{predictive} and \textbf{generative} approaches, mainly for \textbf{cheaper evaluation} and \textbf{better proposal quality}.
Predictive surrogates~\cite{zhao_highly_2023, wang_efficient_2024, mao2023multi} approximate forward solvers for near-instant FoM evaluation inside search loops or for warm-starting in adjoint refinement, but can fail under distribution shift or sparse coverage.
\emph{Generative methods}~\cite{Tang2020GenerativeNanophotonics,Kim2022InverseGANNanophotonics,Ma2019ProbabilisticMetamaterials,Tang2020GenerativeNanophotonics} directly propose geometries conditioned on targets, reducing initialization sensitivity and exploring multi-modal solutions. This is especially valuable for ill-posed, many-to-one inverse problems; Candidates are then followed by physics verification and local, constraint-aware refinement to ensure correctness and manufacturability.

\subsubsection{Applications of photonic inverse-designed devices}
Table~\ref{tab:inv} surveys \textbf{representative inverse-designed devices and their settings} (parameterization, optimizer, design of freedom (DoF), fab-awareness, and the approximate number of electromagnetic simulations (\# Sims)). Inverse design has been applied broadly to \emph{passive} PIC building blocks such as wavelength routers~\cite{wang_efficient_2024} and filters~\cite{mao2023multi}, phase shifters~\cite{liao_inverse_2023}, microring resonators~\cite{gao2022BO}, mode~\cite{liao_inverse_2024} and power splitters~\cite{Tang2020GenerativeNanophotonics, liu_inverse_2025}, grating couplers~\cite{sapra_inverse_2019}, polarization rotators~\cite{yu_genetic-algorithm-optimized_2017}, multimode crossings~\cite{muratsubaki_direct-binary-search_2022}, WDM demultiplexers~\cite{piggott_inverse_2015, schubert_inverse_2022}, and nonlinear optical switches~\cite{hughes_adjoint_2018}.  

Notably, \emph{matrix-vector multiplication (MVM) units}~\cite{wang_inverse-designed_2024, nikkhah_inverse-designed_2024} for ONN and photonic accelerators have emerged as prominent application targets, because compact footprint and engineered spectral responses directly impact scalable, energy-efficient AI hardware.
The paradigm also extends beyond on-chip PICs to fiber, lasers, and free-space optics~\cite{chebaane_machine_2024, li_deep_2023, marques-hueso_genetic_2010}. 
In addition to passive components, inverse design has been demonstrated for \emph{active or tunable} devices, including optical amplifiers~\cite{zhao_highly_2023}, silicon modulators~\cite{zhu_pso-aided_2024}, and PCM-based MMIs~\cite{wu_reconfigurable_2025}, highlighting its relevance to both communication and computing-oriented photonic systems.
  
Across applications, DoF ranges from a few structural parameters to $10^3$--$10^4$ in pixel/free-form geometries, which largely determines the optimizer choice.
Heuristic methods (PSO/GA/DRL/DBS) are common for low-moderate DoF due to better global exploration but higher simulation cost, whereas adjoint-based gradient optimization dominates at high DoF by enabling thousands of variable updates.

Despite these successes, Table~\ref{tab:inv} also highlights a persistent gap between numerical optimality and manufacturable performance. Many pixel/free-form solutions introduce sub-resolution features and sharp geometries that are process-sensitive, motivating \emph{fab-aware} inverse design (FAID)~\cite{Ma2025BOSON, chen2020design, gershnabel2022reparameterization, hammond2022high, NP_khoram2020controlling, wang2019robust, wang2011robust, schevenels2011robust, gershnabel2022reparameterization, hammond2021photonic, chen2020design, mao2023multi}. Existing strategies range from \emph{post-hoc} filtering/regularization, explicit min-feature and deformation constraints, and \emph{in-loop} enforcement (e.g., differentiable lithography or DRC-aware optimization) to restrict search to manufacturable subspaces and improve robustness.

\subsubsection{Prospective and open challenges}
Looking forward, fabrication-aware inverse design must move beyond ``idealized 2D optimization'' toward \emph{system-ready devices under realistic 3D process variation and coupled multiphysics}.
Key directions include:
\textbf{(1) Closing the simulation-fabrication gap with realistic variability models:} capturing 3D effects (sidewalls, roughness, etch/linewidth nonuniformity, index fluctuation) across \emph{local} and \emph{global} scales to translate uncertainty into robust margins.
\textbf{(2) Scalable high-fidelity EM/multiphysics simulation:} 3D EM plus thermal/electrical/mechanical coupling is far costlier than 2D, limiting device size, spectral breadth, and free-form DoF.
\textbf{(3) Fast yet accurate 3D-level AI solvers:} learned surrogates should deliver near-3D fidelity with reliable gradients and calibrated uncertainty for yield-aware optimization; large language model (LLM)-guided orchestration may improve usability~\cite{Kim2025LLMNanophotonics}.
\textbf{(4) Multiphysics-in-the-loop actuation/control:} modeling practical tuning/modulation (heaters) with RC limits, loss, thermal crosstalk, and power to enable robust reconfigurable PICs.
\textbf{(5) Device-circuit co-design and layout-aware constraints:} To make inverse-designed components deployable at the circuit level, optimization must incorporate circuit context and layout constraints, such as routing parasitics, coupling dispersion, thermal proximity, and packaging stress.
\textbf{(6) From device inverse design to circuit-level topology inverse design:}
A natural next step is to extend automated design beyond device synthesis toward \emph{circuit and module optimization}, where device arrangement, interconnection, and parameterization are included in the search space.
Compared to device-level inverse design, circuit-level synthesis introduces a combinatorial discrete search space with complicated constraints, making it substantially more challenging.
Recent work has begun to explore this direction via differentiable and multi-objective optimization to automatically search Pareto-optimal photonic tensor core topologies that improve expressivity, area/energy efficiency, and robustness~\cite{NP_DAC2022_Gu,jiang2025adeptz}.
While still nascent relative to device inverse design, these results suggest a path toward ``design compilers'' for programmable photonic fabrics.

\subsection{EPDA: Circuit/Chip-level Layout Automation}

\begin{figure*}
    \centering
    \includegraphics[width=0.96\textwidth]{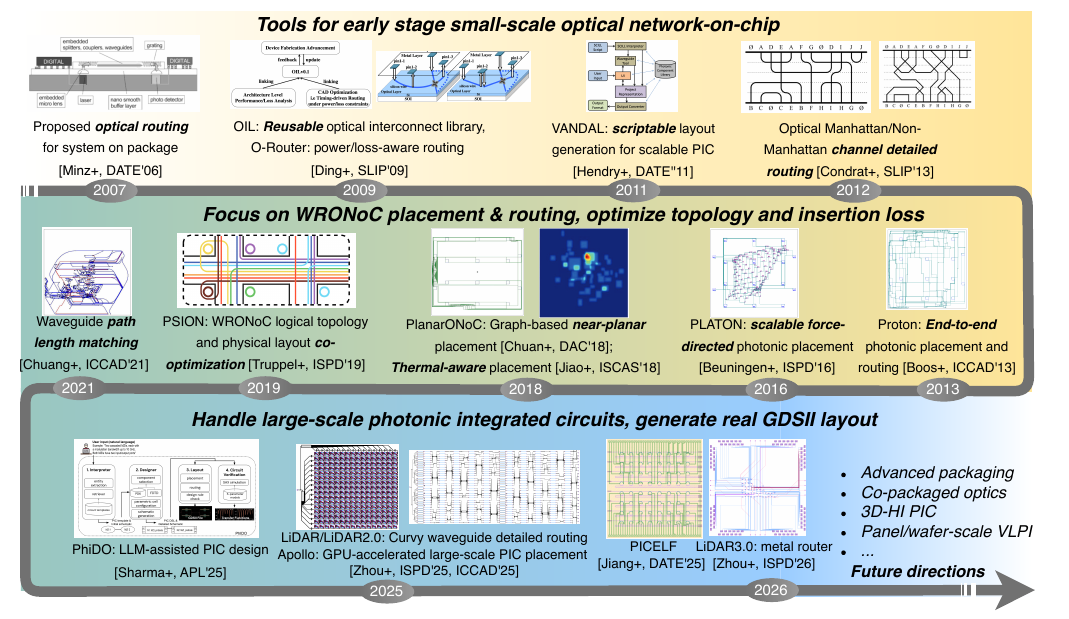}
    \vspace{-5pt}
    \caption{Representative PIC layout automation research milestones and trends.}
    \label{fig:PD_timeline}
    
\end{figure*}

After obtaining the PIC netlist and component layouts, designers often use schematic-driven layout (SDL)~\cite{bogaerts2018silicon}: manually place components and connect them with waveguides and wires.
As PICs scale to AI systems with many components, this manual loop becomes a bottleneck, labor-intensive, error-prone, and hard to iterate since small schematic changes trigger re-placement/re-routing and repeated design-rule checks (DRC).
This motivates PIC layout automation for faster iteration, scalability, and improved quality.

\subsubsection{Challenges of PIC layout automation}
\noindent \textbf{(1) Layout Sensitivity and Physics-Aware Constraints.}
PIC layout is \emph{inherently} performance-driven: the geometry directly dictates system behavior. 
A simple waveguide can be a functional element whose length sets phase and whose curvature/crossings set loss and crosstalk. And placement/routing must enforce constraints such as thermal crosstalk mitigation during heater placement, path-length matching for MZMs, and cross-domain exclusions (e.g., avoiding long metal overlap that increases optical loss).
\noindent \textbf{(2) Resource-Limited Physical Layout.}
Large-scale EPICs are constrained by area and limited routing layers (often a single optical layer and a few metal layers), so routability depends on early corridor planning and minimizing crossings/detours, including minimizing waveguide crossings, reducing via usage and detours in metal routing.
And chip packaging further pins optical I/O and pads to fixed locations, creating rigid boundary conditions that concentrate congestion and make routing topologically complex.
\noindent \textbf{(3) High-Speed Circuit Layout Challenges.}
High-speed links add transmission-line constraints (impedance/group-index matching), requiring area-hungry geometries (e.g., coplanar waveguides), careful handling of bends and metal fill, and shielding for differential pairs.
\noindent \textbf{(4) Scalability and Fabrication Metrics.}
At the thousand-component scale, automation is essential, but must account for yield (process variation, lithography impacts on $n_{\text{eff}}$).
The goal is to balance area, insertion loss, and electrical bandwidth under manufacturing and physical constraints.

\subsubsection{Tools for PIC layout automation}
As summarized in Fig.~\ref{fig:PD_timeline}, we categorize prior research on PIC
physical design automation into three stages.

\textbf{(i) Early-stage, small-scale ONoC-driven tools (2007--2012).}
Motivated by \emph{optical networks-on-chip} (ONoCs), early works emphasized \emph{waveguide routing} and lightweight automation, including timing/congestion-aware optical routing for 3D system-on-packag~\cite{minz_optical_2007}, reusable parameterized libraries (OIL)~\cite{ding_oil_2009}, power-aware routing optimization (O-Router)~\cite{ding_o-router_2009}, and scriptable layout generation (VANDAL)~\cite{hendry2011vandal}.
In parallel, channel-level Manhattan/non-Manhattan detailed routing was also explored to better capture geometry and crossings in constrained regions~\cite{condrat2012methodology, sharma2023optimizing, condrat2013channel}.

\textbf{(ii) WRONoC placement-and-routing with topology awareness (2013--2021).}
Research expanded toward automated placement and routing (P\&R) for wavelength-routed ONoCs (WRONoCs), with increasing \emph{co-optimization} between logical topology and physical layout.
PROTON~\cite{boos_proton_2013} pioneered end-to-end photonic P\&R by combining nonlinear placement with Lee-style routing, while PLATON~\cite{von_beuningen_platon_2016} introduced scalable force-directed placement.
To reduce crossings and improve routability, PlanarONoC used planar/graph-based reasoning~\cite{chuang_planaronoc_2018}, and later works pursued topology--layout co-optimization~\cite{chen_cponoc_2025}. Complementary studies also addressed key layout tasks such as path-length matching~\cite{chuang_-chip_2021}, and structure-aware routing to reduce detours/crossings~\cite{zheng_topro_2021}.
In parallel, the community also advanced the \emph{schematic-driven} methodology~\cite{chrostowski2016schematic, chrostowski2016design}: the flow starts from a circuit schematic, uses a PDK-backed component library for immediate simulation, and then generates layout based on the schematic, followed by DRC verification and post-layout parameter extraction to update the schematic.
This paradigm is also widely adopted in industry; commercial toolkits (e.g., Synopsys OptoCompiler and Cadence Virtuoso) and open-source frameworks (e.g., GDSFactory) support GUI-, API-driven layout generation, guided routing, and exporting netlists for simulation/verification~\cite{GDSFactory}.

\textbf{(iii) Tackling emerging large-scale PIC design automation (2022--present).}
With photonic computing pushing denser circuits, focus shifts to \emph{circuit-scale} automation that outputs manufacturable GDSII.
LiDAR/LiDAR2.0~\cite{LiDAR_ISPD_Zhou, zhou2025lidar} advances detailed routing via dynamic crossing insertion and curvilinear routing, producing near-DRV-free final layout on WRONoC and photonic-computing designs.
In parallel, the work~\cite{wu2025automatic} proposes an optical routing flow targeting phase/delay matching, using diffusion-based length matching with spiral detours.
To further reduce loss under richer process options,  the work~\cite{wu2025constraints} explicitly models hybrid waveguides and transitions, and optimizes insertion loss while enforcing matching constraints.
As electrical nets scale, metal routing becomes a bottleneck~\cite{jiang2025picelf}; to address photonics-specific keep-outs and spacing, recent work proposes congestion/DRC-aware global electrical planning with waveguide-aware assignment and guidance-driven detailed routing~\cite{zhou2025photonics}. In addition, PICELF~\cite{jiang2025picelf} targets the electronic routing by assigning the electrical pin via nonlinear binary programming and performing a fast two-stage router to produce DRC-clean metal layouts.
Besides routing, Apollo~\cite{PLACE_ICCAD2025_Zhou} introduces GPU-accelerated, routing-informed placement with bending-aware objectives and explicit congestion/crossing modeling, achieving a $94.79\%$ routing success rate on large-scale photonic-computing benchmarks.
Beyond classical algorithmic EDA, emerging \emph{agentic} workflows explore natural-language-to-GDSII automation. The PhIDO multi-agent framework~\cite{sharma2025ai} demonstrates an end-to-end pipeline that translates natural-language PIC requests into structurally valid layouts.

\subsubsection{Prospective and open challenges}
Looking forward, photonic layout automation is expected to evolve in response to the emerging paradigm of \textbf{very-large photonic integration (VLPI)}, where hundreds to thousands of photonic devices are tightly integrated with electronics to form heterogeneous, programmable systems.
EPDA must support \textbf{heterogeneous EPIC design} under advanced packaging and system-level constraints.
First, \textbf{packaging- and co-packaged optics (CPO)-driven design} will become a central requirement: co-packaged optics introduces tight constraints on bump/through silicon via (TSV) placement, micro-assembly, fiber/laser coupling interfaces, thermal management, and power delivery.
EPDA tools must therefore enable joint optimization across die, interposer, and package hierarchies, balancing optical loss, electrical signal integrity, thermal density, and manufacturability within a unified design space.
Second, \textbf{3D PIC and multi-layer photonics} represent a major opportunity and challenge. Expanding beyond single-layer silicon waveguides to multi-layer and truly 3D interconnects can unlock unprecedented integration density, but it also requires new placement/routing abstractions (layer assignment, 3D waveguide vias, vertical couplers), 3D-aware design rules, and cross-layer crosstalk/thermal constraints that current 2D routing-centric workflows cannot capture. 
Third, \textbf{robustness and first-pass manufacturability} will increasingly rely on \textbf{yield- and variability-aware optimization}. Incorporating process variations (linewidth/etch bias, thickness, overlay) and variability models directly into placement, routing, and device selection can reduce late-stage iterations and improve first-pass success. In this context, \textbf{machine-learning-guided heuristics} offer a promising direction to accelerate exploration and provide better initial solutions (e.g., routability-aware placement seeds, rapid loss/crosstalk estimators), while still requiring physics-based refinement and guarantees. 
Finally, scalable VLPI-based AI system deployment demands \textbf{verification and signoff beyond geometric DRC}: in addition to geometric rule checking, future EPDA flows must support reliable photonic layout-versus-schematic (LVS) (device recognition and parameter extraction), proximity- and layout-dependent effect checking (e.g., waveguide coupling/crosstalk, crossing/bend penalties), and consistent post-layout back-annotation for system-level co-simulation across optical, electrical (RF), and thermal domains.

Together, these EPDA capabilities are essential to translate future VLPI designs from layout to first-pass hardware with predictable system behavior, enabling photonic systems to scale with the same rigor that electronic ICs achieved.

\section{Conclusion and Outlook}
\label{sec:future}
As machine intelligence becomes a pervasive infrastructure layer, compute demand is outpacing the energy and bandwidth gains of post-Moore silicon.
This review has argued that photonic machine intelligence is entering a new phase, shifting from demonstrating physical feasibility to establishing system-level scalability and reproducible advantage.
Realizing the potential of photonic computing requires a fundamental change in design philosophy. 
We conclude that the future of photonics lies not only in device foundation optimization, but also in a holistic system co-design where optical physics, electronic interfaces, and AI workloads are jointly optimized to create new system degrees of freedom.
A primary insight of our analysis is that hardware must co-evolve with the rapid advancements in AI algorithms. 
To move beyond narrow, static inference, photonic AI systems must prioritize workload-flexible programmability and consistent computing fidelity, sustaining versatility and accuracy for dynamic, evolving workloads.
This motivates a closed feedback loop between software and hardware, where device, circuit, system, algorithm, and physical implementation are designed together.

Sustaining this trajectory will require design workflows that scale with complexity.
We identify EPDA as the pivotal enabler and new research focus for the community. 
The field must transition to a full-lifecycle design ecosystem that integrates (i) AI-assisted simulation and inverse design for compact, manufacturable components, (ii) rigorous cross-layer modeling for fair benchmarking and bottleneck attribution, and (iii) automated, verification-aware physical design for large-scale heterogeneous integration.

\noindent\textbf{Outlook and Final Remark.}~Key directions for the field include:
\ding{202}~\uline{Standardized benchmarking and system evaluation}.~
Progress will increasingly hinge on community benchmarks that are physically rigorous (grounded in realistic device/interface models), system-comprehensive (including conversion, control, memory, interconnect), and workload-aware (exploring accuracy-efficiency tradeoffs on diverse applications and mapping strategies), enabling fair comparisons.
\ding{203}~\uline{Cross-layer co-design with versatility and robustness as key focus}.~
To remain relevant amid rapid algorithmic change, photonic AI must move from narrow, fixed-function demonstrations toward workload-flexible platforms, where robustness is treated as a first-class design constraint and addressed end-to-end, from device physics and mixed-signal interfaces to system control and algorithm-aware calibration/adaptation that sustains consistent accuracy over time.
\ding{204}~\uline{Open, reusable, full-stack EPDA infrastructure}.~
A decisive accelerant will be a full-stack EPDA toolflow, spanning simulation, inverse design, system modeling, and automated physical layout, that expedites design cycles, improves resilience and yield, and unlocks new design degrees of freedom by leveraging modern AI-driven methodologies and high-performance computing, ultimately turning lab-scale prototypes into reproducible ecosystems.

\section*{Acknowledgment}
This work is supported in part by the AFOSR Multidisciplinary University Research Initiative (FA9550-17-1-0071), Air Force Office of Scientific Research (FA9550-23-1-0452) on Photonics for AI and AI for Photonics, Texas Center for Optical Computing and Interconnects, and equipment donations from Nvidia.


\vfill

\end{document}